\documentclass[10pt,a4paper]{article}

\usepackage{amsmath,refcount}
\usepackage[latin1]{inputenc}
\usepackage{fancyref}
\usepackage{authblk}
\usepackage{amsmath}
\usepackage{amsfonts}
\usepackage{amssymb}
\usepackage{graphicx}
\usepackage[svgnames]{xcolor}
\usepackage{tikz}
\usepackage{url}

\usetikzlibrary{decorations.markings}
\usetikzlibrary{shapes.geometric}
\usepackage{tkz-graph}
\usepackage{hyperref}
\usepackage[caption = false]{subfig}

\usepackage{graphicx}
\usepackage{multicol}
\usepackage{footnote}
 
\pgfdeclarelayer{edgelayer}
\pgfdeclarelayer{nodelayer}
\pgfsetlayers{edgelayer,nodelayer,main}

\usepackage{float}

\tikzstyle{none}=[inner sep=0pt]

\tikzstyle{rn}=[circle,fill=Red,draw=Black,line width=0.8 pt]
\tikzstyle{gn}=[circle,fill=Lime,draw=Black,line width=0.8 pt]
\tikzstyle{yn}=[circle,fill=Yellow,draw=Black,line width=0.8 pt]

\tikzstyle{simple}=[-,draw=Black,line width=2.000]
\tikzstyle{arrow}=[-,draw=Black,postaction={decorate},decoration={markings,mark=at position .5 with {\arrow{>}}},line width=2.000]
\tikzstyle{tick}=[-,draw=Black,postaction={decorate},decoration={markings,mark=at position .5 with {\draw (0,-0.1) -- (0,0.1);}},line width=2.000]

\tikzstyle{newstyle}=[
-,draw=Green,line width=2.000]
\usepackage{cprotect}
\usepackage{algorithm}
\usepackage[noend]{algpseudocode}
\usetikzlibrary{decorations.markings}
\usetikzlibrary{shapes.geometric}
\makeatletter
\def\BState{\State\hskip-\ALG@thistlm}
\makeatother

\author[1,4]{Marcos Matabuena}

\author[3]{Carlos Meijide-García}
\author[2]{Pablo Rodríguez-Mier}
\author[1]{Víctor Leborán}
\affil[1]{CiTIUS (Centro Singular de Investigación en Tecnoloxías Intelixentes), Universidade de Santiago of Compostela,
	Santiago de Compostela, Spain}
\affil[2]{Toxalim (Research Centre in Food Toxicology), Université de Toulouse, INRAE, ENVT, INP-Purpan, UPS, Toulouse, France}
\affil[3]{Universidade de Santiago of Compostela,
	Santiago de Compostela, Spain}

\affil[4]{\url{marcos.matabuena@usc.es}}

\title{COVID-19: Estimating  spread in Spain solving an inverse problem with a probabilistic model}

\begin{document}

	\maketitle

	\section*{Abstract}
	We introduce a new probabilistic model to estimate the real spread of the novel SARS-CoV-2 virus along regions or countries.   Our model simulates the behavior of each individual in a population according to a probabilistic model  through an inverse problem; we estimate the real number of recovered and infected people using mortality records. In addition, the model is dynamic in the sense that it takes into account the policy measures introduced when we solve the inverse problem. The results obtained in Spain have particular practical relevance: the number of infected individuals can be $17$ times higher than the data provided by the Spanish government on April $26$ $th$ in the worst case scenario. Assuming that the number of fatalities reflected in the statistics is correct, $9.8$ percent of the population may be contaminated or have already been recovered from the virus in Madrid, one of the most affected regions in Spain. However, if we assume that the number of fatalities is twice as high as the official numbers, the number of infections could have reached $19.5$\%.  In Galicia, one of the regions where the effect has been the least, the number of infections does not reach $2.5$ \%. Based on our findings, we can: i) estimate the risk of a new outbreak before Autumn if we lift the quarantine; ii) may know the degree of immunization of the population in each region; and iii) forecast or simulate the effect of the policies to be introduced in the future based on the number of infected or recovered individuals in the population.

	\section{Introduction}

	The spread of the Coronavirus is generating an unprecedented crisis worldwide. In Europe, for example, it is considered to be the most significant challenge that the continent has faced since  World War II. In the light of this situation of emergency, the governments must work quickly to avoid the collapse of the healthcare system, reduce the mortality associated with the virus, and avoid the possible effects of an economic recession \cite{Kickbuschm406,Layneeabb1469}.

	Given the strong capacity of the virus to spread and the lack of preventive measures, many countries have been systematically forced to lockdown the population temporarily. Although these policies may be useful in controlling the spread of the virus in the immediate future, they are unsustainable over time from an economic point of view. In this regard, forecasting the evolution and consequences of the pandemic based on the exposure of the population becomes a critical factor in decision-making \cite{enserink2020mathematics,kissler2020projecting}. However, to rigorously predict these effects, it is necessary to assess the current spread of the epidemic, which is often unknown.

	At the beginning of the XX century, the first mathematical models to study the dynamics of an epidemic were introduced. Probably the most well-known is the susceptible-infected-recovered model (SIR). SIR  models and its variations \cite{Huppert2013,keeling2011modeling} divide the population into compartments, and using differential (deterministic) equations, the number of individuals in each of the compartments over time is estimated. Since then, many new variations of these models have been introduced in the literature (see for review \cite{mandal2011mathematical,allen1994some,satsuma2004extending,ball2019stochastic} or more contemporary examples \cite{choi2012inference,osthus2017forecasting}).

     However,  those approaches suffer from a critical shortcoming. In essence, they focus on explaining the dynamics of the epidemic at the population level and exclude the complex interactions that occur at the microscopic level between individuals \cite{Koopman1999,Gomez-Gardenes2016,Keeling2005}. Furthermore, these models tend to be adjusted with fixed parameters that are not updated over time, according to how the epidemic evolves and the different policies introduced. 
	
	 In the current era of precision medicine, where more efficient solutions are being sought that optimize the health of the individual \cite{kosorok2019precision}, it is surprising not to find in the current state of the art models of epidemic spread that follow more individual approaches \cite{Reich3146,10.1371/journal.pcbi.1006134}, as is rightly pointed out in \cite{Viboud2802}. Precisely in \cite{Viboud2802}, the authors propose to adopt new models that exploit the enormous individual information recorded by biosensors \cite{Li2017} and other devices.

	A crucial aspect to consider in the success of mathematical modeling is the need to adjust the parameters of the models with accurate data. However, this is difficult to achieve in practice, because governments are often overloaded, and information is only recorded on patients with symptoms. The observational nature of this data \cite{greenland2005multiple}, together with a delay or errors in the recording of information \cite{hernan}, requires specific techniques to correct the biases related to the data \cite{biometrics}.

	\subsection{Main contributions and results}
	
	Motivated by the need to developed new models that overcome these drawbacks and that estimate the expansion of the SARS-CoV-2 in a realistic way in Spain, we introduce a new probabilistic model. The contributions and most significant results of this paper are introduced below.

	\begin{itemize}
		\item    To the best of our knowledge, we propose one of the few probabilistic models that estimate the evolution of an epidemic from each individual of the population. This model combines the finite Markov model, along with continuous probability distributions and a dynamic Poisson model.

		\item The model is designed in such a way that it does not need to use statistics on the number of infections that have a significant measurement error, and uses mortality records instead.
		
		\item The model estimates for each day the number of susceptible, contaminated, recovered, and fatalities. For this, we solve an inverse problem, and we introduce biological expert knowledge related to the SARS-CoV-2 disease into the probabilistic model about the time each individual spends in the different states and the transition of probabilities.

		\item  Using the data from several regions of Spain, we estimate the real number of infections as of April $26$st. The results show a higher proportion of infections than that provided by the Spanish government.

		\item The effect of the Coronavirus has been uneven among regions of Spain. In Madrid, more than $10$  percent of the population was infected, or it recovers from the virus, while in Galicia, this percentage is less than $2.5$.
		
		\item      These results highlight differences between regions; i) the degree of immunity of the population can be different; ii) the number of recovered people who can return to normality is also different; iii) based on this analysis, a reasonable strategy to be considered could be to lift the quarantine  on a case-by-case basis based  on the degree of exposure of each region to the virus.

		\item  Finally, we repeat the estimations assuming that the number of fatalities is twice  than reported by the official records. In this case, the number of infections can increases between $40$ y $110$ percent by region.  However, we must interpret these results with caution.
		
	\end{itemize} 
	
	\subsection{Outline}
	
	The mathematical details of the model are relegated to the end of the article, to facilitate the readability  for a general target audience. The structure of the article is as follows: First, we describe the evolution of Coronavirus in Spain, together with some demographic and economic characteristics of the Spanish population.  Subsequently, we fit our model in different scenarios, and we show the evolution of susceptible patients, infected and recovered ones over time.  Next, we discuss the results and limitations of the model. Finally, we introduce the mathematical content of our model and the computational details of the implementation and parameter optimization.

	\section{COVID-19 in Spain}

	Spain was one of the first countries in the world to experience the effects of COVID-19 after China and Italy. However, despite the delay in the start of the outbreak, with respect to these countries, the consequences are now more dramatic. To give a better context to the evolution of Coronavirus in Spain and to compare it with other countries, we introduce some historical background:

	\begin{itemize}
		
		\item January $31$st. The first positive result was confirmed on Spanish territory in La Gomera. At that time, there were around $10,000$ confirmed cases worldwide.
		
		\item February $12$th. The Mobile World Congress, one of the most remarkable technological congresses in the world to be held in Barcelona, was canceled.

		\item March $8$th.    Multitudinous marches were celebrated in Spain. Also, sports competitions and other events were held as usual.  
		
		\item March $13$th. Madrid reported $500$ new cases of Cov-19 in one day ($64$ deaths total). Wuhan had gone into lockdown with $400$ new cases per day ($17$ deaths total). 
		
		\item March $14$th. With the increase in the outbreak of infections, the government declared a quarantine throughout the country.
		
		\item March $21$st. Due to an overloaded health system, the first patients started to arrive at new makeshift hospitals.
		
		\item April $3$rd. Spain accounts for a total of $117,710$ confirmed cases, surpassing Italy for the first time.
		
		\item April $6$th. Spain becomes the country in the world with more deaths per million inhabitants. 
		
		\item April $9$th. The FMI forecasts that $170$ countries are going to be into recession this year in the worst crisis since the Great Depression.
		
		\item April $18$th. The Spanish government changes protocols for the daily statistics of COVID-19.
		
		\item April $21$th. The president of the United States, Donald Trump, declared: "It's incredible what happened to Spain, it's been shattered.".

	\end{itemize}

\begin{figure}[ht!]
\centering
		\scalebox{0.85}{\includegraphics[width=1.3\linewidth]{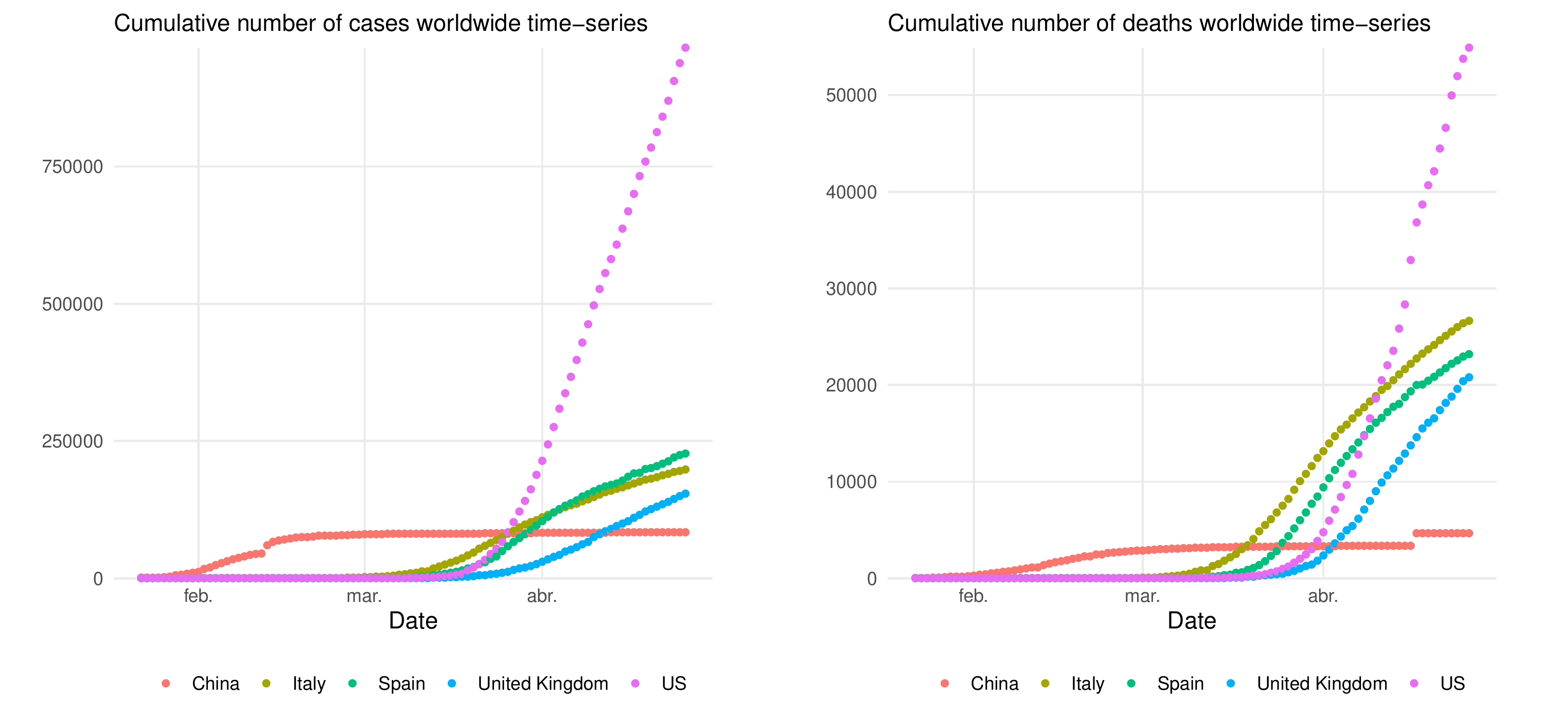}}
		\caption{Spread and number of deaths of Coronavirus in Spain, Italy, China, and the United States. Number of accumulated infected patients (left) and the number of accumulated deaths (right) \cite{w1}. }
		
				\label{fig:world}
	\end{figure}

	 Figure \ref{fig:world}, shows  accumulated number of cases and deceases respectively in the previous periods in Spain, Italy, China, United Kingdom and the United States according to the data supplied by the different governments.
	
	\begin{table}[ht!]
		\centering
		\scalebox{0.8}{
			\begin{tabular}{crrrrrr}
				\hline
				& Galicia & País Vasco & Castilla y León & Cataluña & Madrid & España \\ 
				\hline
				Population & $2,698,763$  & $2,181,916$  &  $2,553,301$  &   $7,609,497$ & $6,685,470$ & $47,100,396$  \\

				At-risk-of-poverty rate & $18.8$ & $8.6$ & $16.1$ & $13.6$ & $16.1$ & $21.5$ \\
				
				Population density  & $91.28$ & $305.19$ & $25.47$& $239.01$ & $830.02$ & $93.08$ \\
				\hline
				\multicolumn{7}{c}{Percentage of population by age group} \\
				\hline
				$0-9$ & $7.50$ & $8.93$ & $7.51$ & $9.78$ & $9.92$ & $9.28$ \\ 
				$9-18$ & $7.56$ &$ 8.78$ & $7.79$ & $9.74$ & $9.44$ & $9.37$ \\ 
				$18-30$ & $10.39$ & $10.66$ & $10.67$ & $12.72$ & $12.80$ & $12.42$ \\ 
				$30-45$ & $21.18$ & $20.28$ & $19.47$ & $22.24$ & $23.24$ & $22.07$ \\ 
				$45-60$ & $22.77$ & $23.28$ & $23.60$ & $21.77$ & $22.23$ & $22.61$ \\ 
				$60-80$ & $22.60$ & $21.41$ & $22.30$ & $18.24$ & $17.33$ & $18.70$ \\ 
				from $81$ on & $8.01$ & $6.67$ & $8.65$ & $5.50$ & $5.03$ & $5.56$ \\ 
				\hline
		\end{tabular}}

		\caption{Demographic  and socioeconomics characteristics  of the Spanish population throughout  some regions:  Galicia, Pa\'{i}s Vasco, Castilla y Le\'{o}n, Cataluña, Madrid \cite{w2}. }  
				\label{table:regiones}
	\end{table}

	Following the statistics of the Population Reference Bureau, Spain is the 20th country with the world's oldest population \cite{w3}. The country demographic structure, poverty rates, and epidemiological profiles is essential to compare mortality between countries.  In the Coronavirus disease, relative and absolute case-fatality risk (CFR) \cite{ghani2005methods} increases dramatically with age and with comorbidity, as evidenced by the current literature. Relative risk can increase by more than $900$\% in patients over $60$ \cite{Zhao2020.03.17.20037572}.
	
	Subsequently, we perform a  descriptive analysis in the regions of Spain that we analyze in this paper: Galicia, Pa\'{i}s Vasco, Castilla y Le\'{o}n, Madrid, Cataluña. Table  \ref{table:regiones}   contains the essential demographic and socioeconomic characteristics of these regions. We can see that Castilla y Le\'{o}n is the region with the highest proportion of elderly people. At the same time, Castilla y Le\'{o}n has the most delocalized population centers, and the Pa\'{i}s Vasco is the region with the lowest poverty rate.  Spain is a multicultural country where there are significant economic, geographical, social, and demographic differences throughout the regions. All these peculiarities make Spain  an interesting country to extrapolate the effects of the spread of the Coronavirus to other regions and countries.

	\begin{figure}[ht!]
	\centering		\scalebox{0.80}{\includegraphics[width=1.3\linewidth]{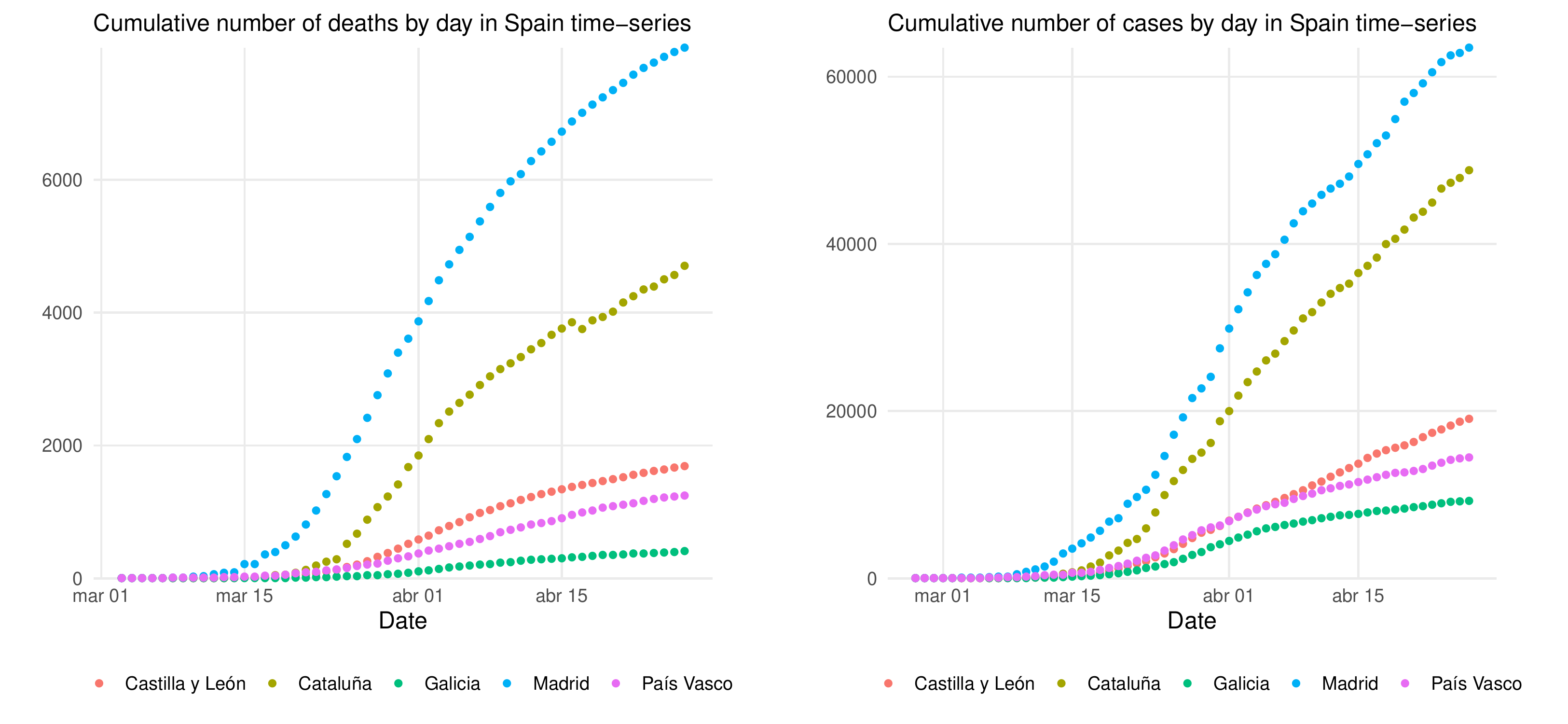}}
		\caption{Evolution of accumulated infected (left) and death patients (right) in Galicia, Pa\'{i}s Vasco, Castilla y Le\'{o}n, Madrid, Cataluña}
		\label{figure:comunidades}
	\end{figure}

Finally, in Figure \ref{figure:comunidades}, we show the evolution of infections and fatalities among the regions under consideration. As we can see, Madrid is the most affected region, while Galicia is the least affected, despite its older population. However, it is important to note that the outbreak began later, and the containment was carried out earlier than in Madrid.

	\section{Results}

	Next, we show the estimation performed with the model until April $26$st in  Galicia, Pa\'{i}s Vasco, Castilla y Le\'{o}n, Madrid, Cataluña,  assuming that these two scenarios hold:
	
	\begin{enumerate}
		\item We assume that the number of real deaths due to Coronavirus is the one reflected by the official records.
		\item We assume that many people have died of Coronavirus, but they have not been included in the records because a diagnostic test was not performed. In particular, we shall suppose that the number of deaths is twice as high as those indicated in the official records each day.
	\end{enumerate}

	We graphically represent along time the result of estimating how many individuals belong to each of the following states:
	
	\begin{itemize}
		\item  $I_1(t)$: Number of infected individuals who are incubating the virus on day $t$.
		\item  $I_2(t)$:     Number of infected people who have passed the incubation period and do not show symptoms on day $t$.
		\item  $I_3(t)$:  Number of infected people who have passed the incubation period and do show symptoms on day $t$.
		\item  $R_1(t)$: Number of recovered cases which are still  able to infect on day $t$.
		\item  $R_2(t)$. Number of recovered cases which are not able to infect anymore on day $t$.
		\item  $M(t)$: Number of deaths on day $t$.      
	\end{itemize}

	In addition, to understand the meaning of the results, we show: i) the number of people who may be contaminated or who have already transmitted the virus as a percentage of the population size; and ii) the rate of new infections each day.

	Finally, we introduce confidence bands of our estimations \footnote{Our confidence bands shall not be confused with classic frequentist statistics confidence bands}.
	\subsection{Scenario 1}
	
	\subsubsection*{Galicia}
	\begin{figure}[H]
		\centering
		\includegraphics[width=1\linewidth]{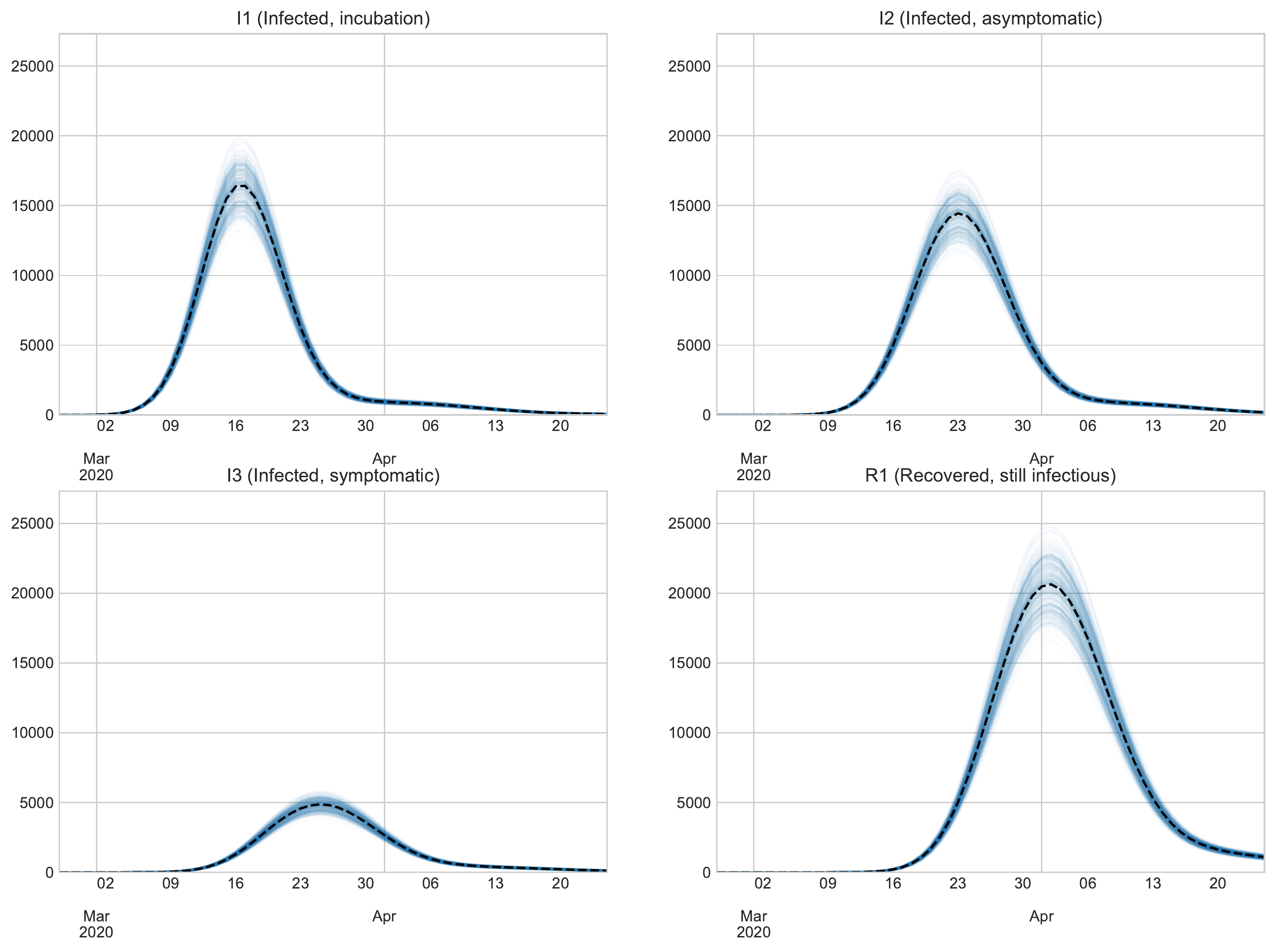}
		\includegraphics[width=1\linewidth]{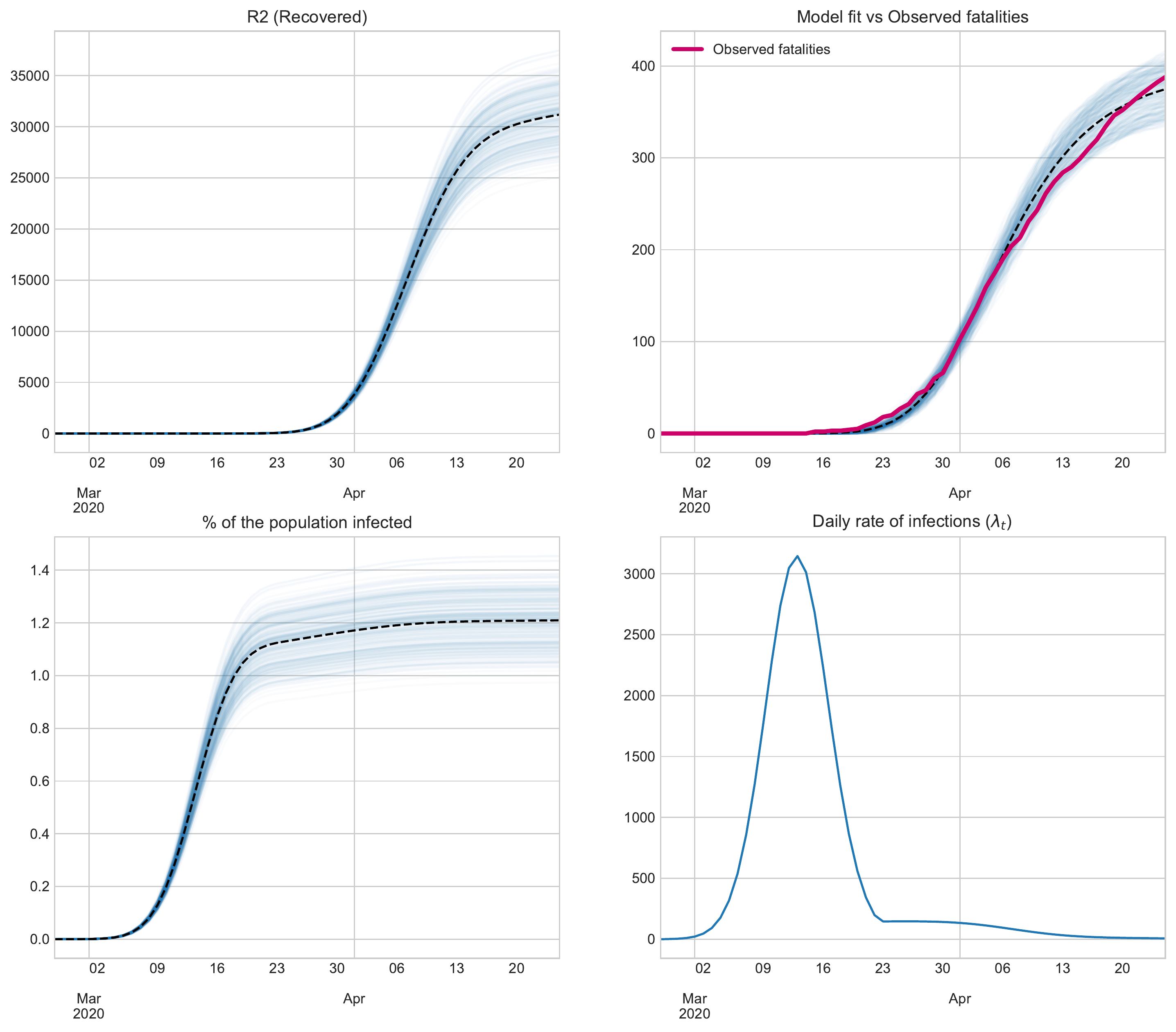}

		\caption{Results in Galicia: we assume that there are as many fatalities as in the official records.}
		\label{fig:fig1gal}
	\end{figure}

%
	
	\subsubsection*{Castilla y León}

	\begin{figure}[H]
		\centering
		\includegraphics[width=1\linewidth,scale=5]{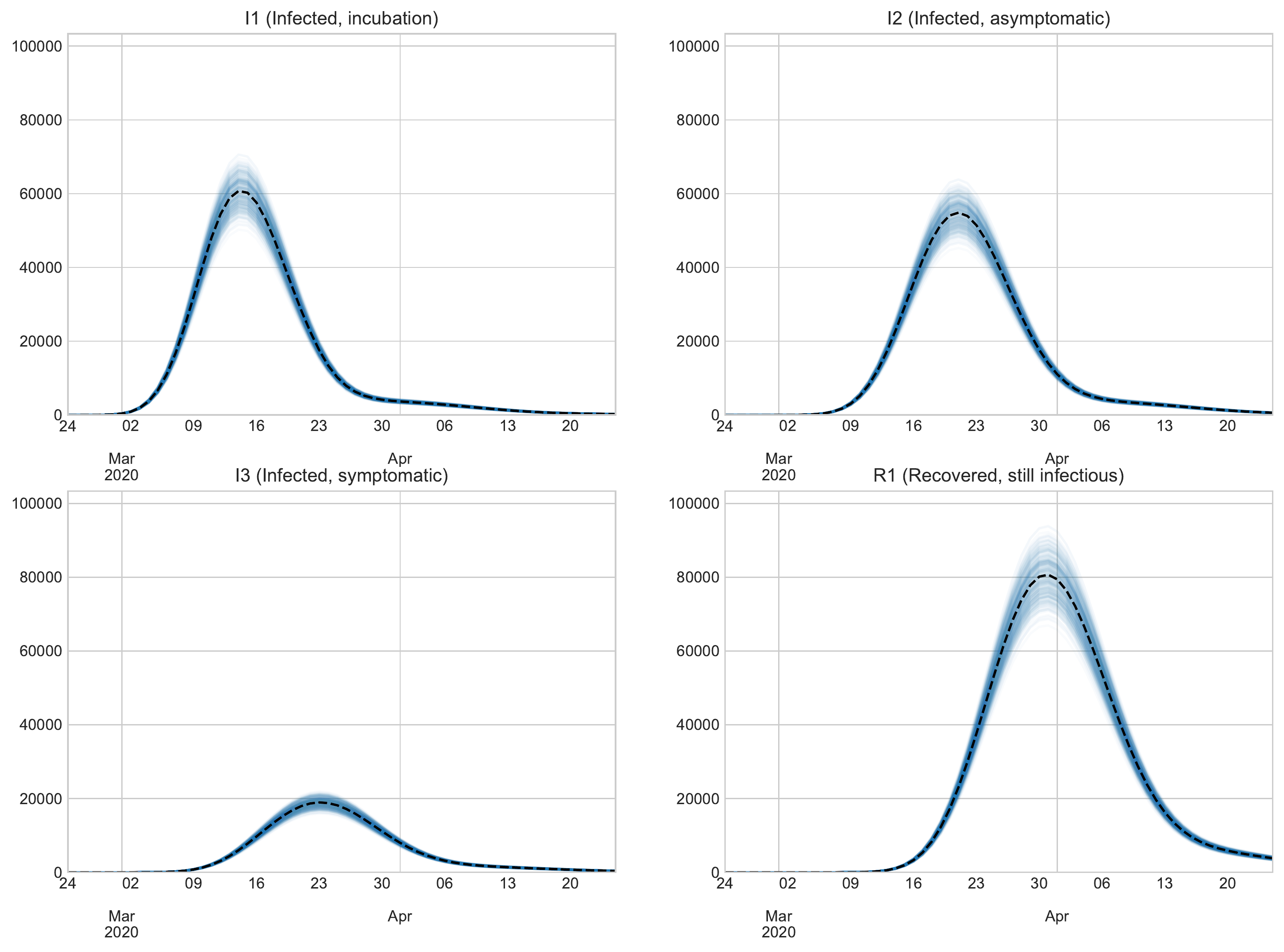}
	    \includegraphics[width=1\linewidth,scale=5]{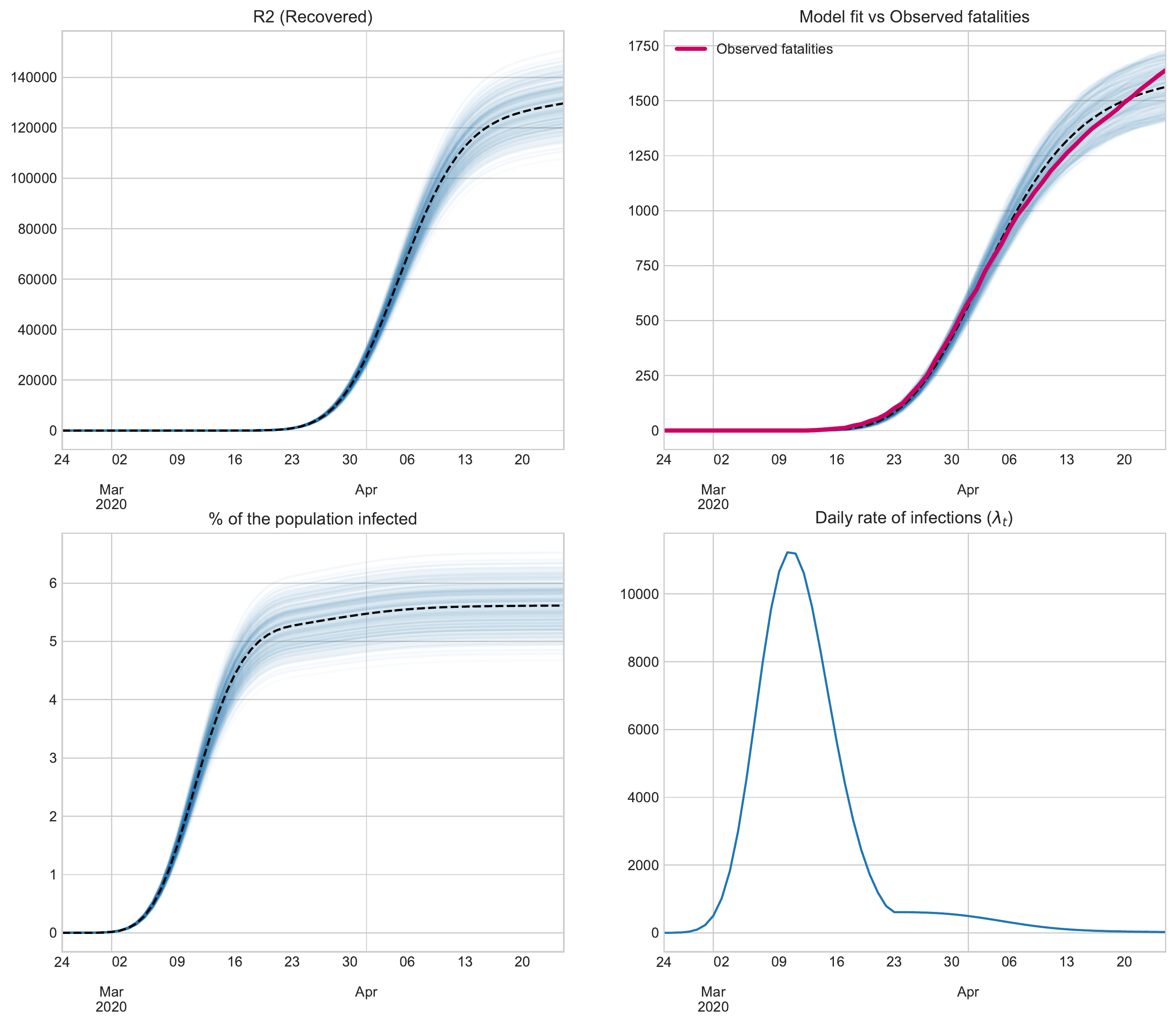}
		\caption{Results in Castilla y León: we assume that there are as many fatalities as in the official records.}
		\label{fig:fig1gal}
	\end{figure}

%
%
%
%
%
	\subsubsection*{País Vasco}

		\begin{figure}[H]
		\centering
		\includegraphics[width=1\linewidth,scale=5]{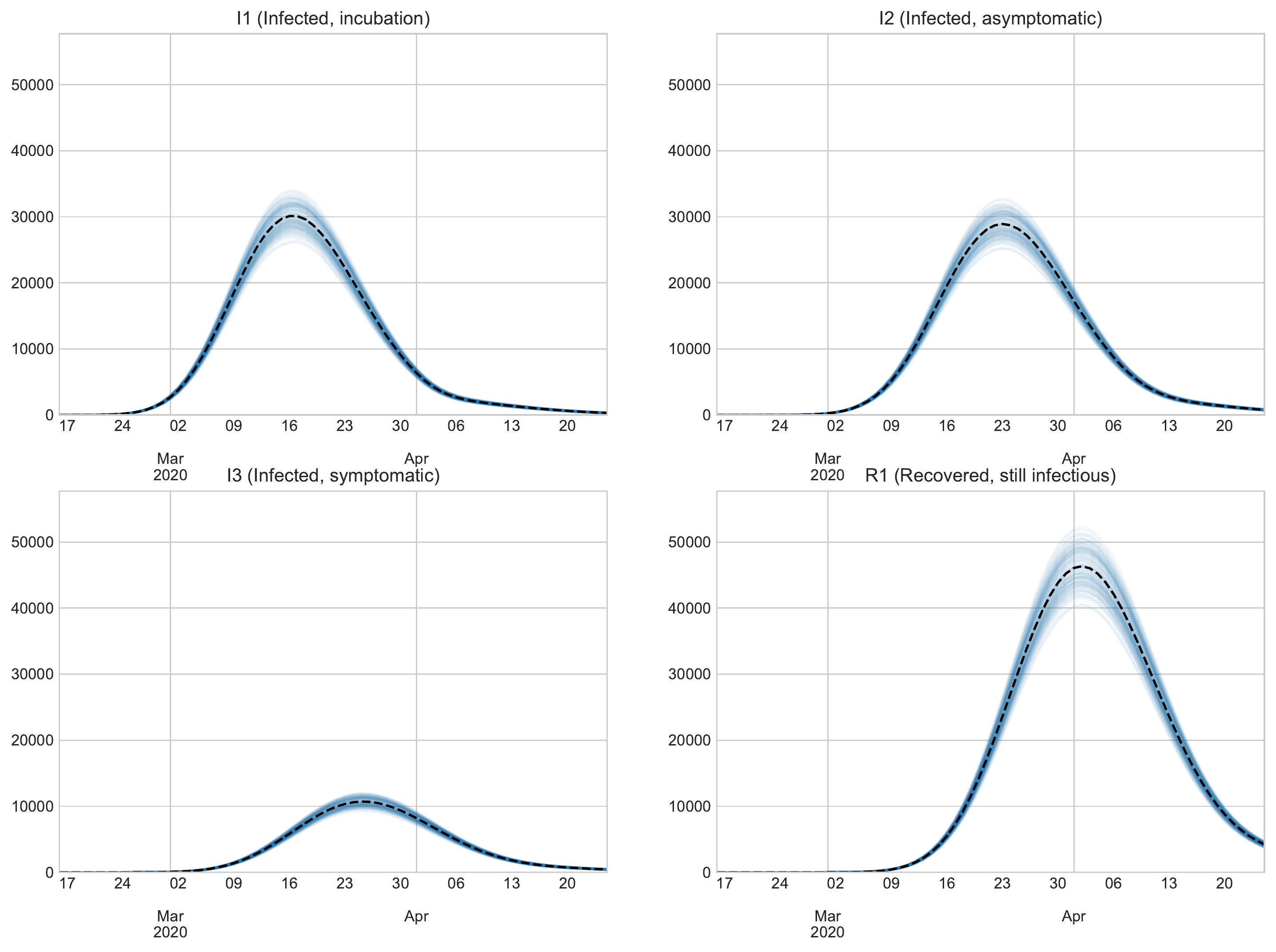}
				\includegraphics[width=1\linewidth,scale=5]{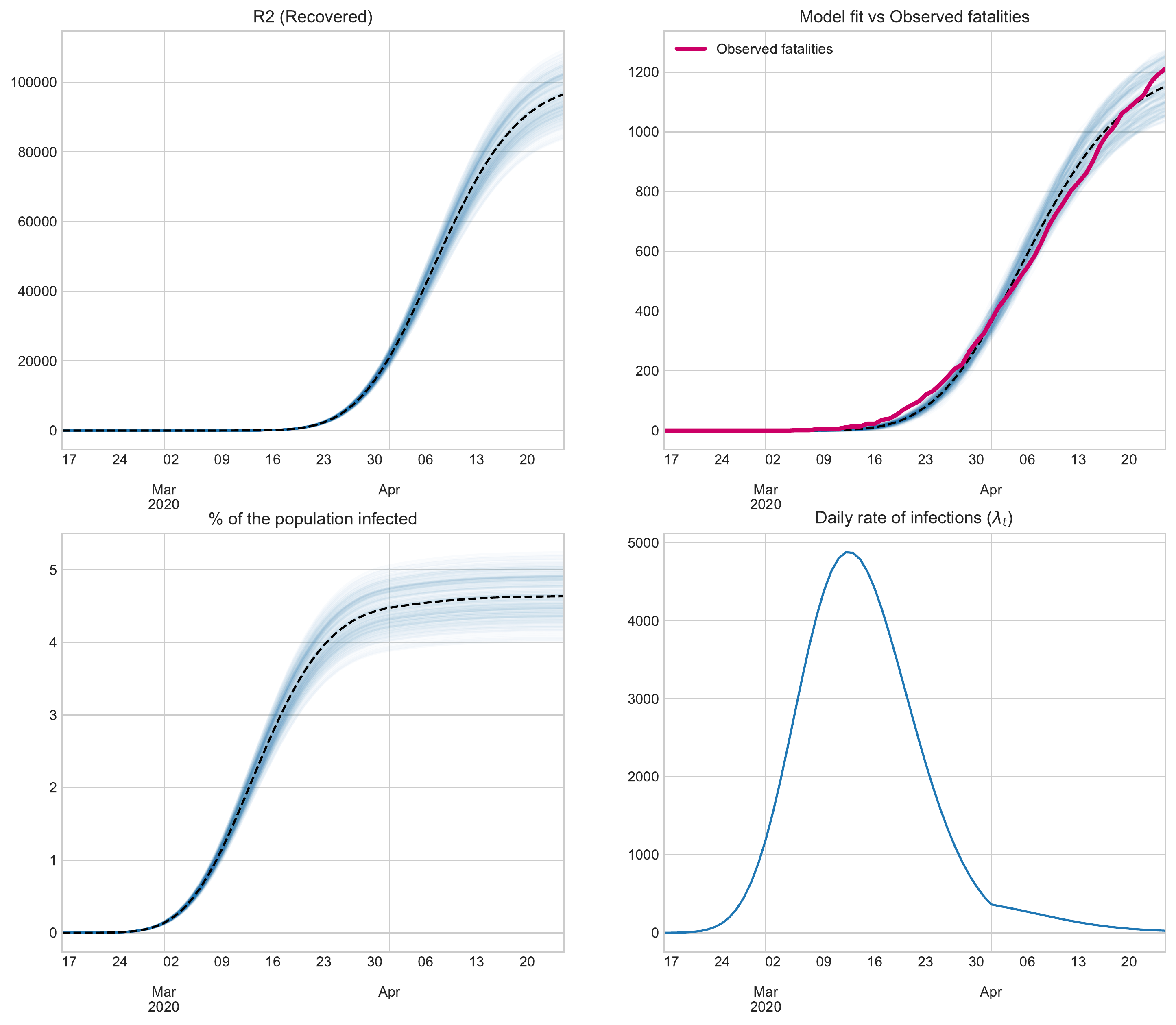}
		\caption{Results in País Vasco: we assume that there are as many fatalities as in the official records.}
		\label{fig:fig1gal}
	\end{figure}

	\subsubsection*{Madrid}

	\begin{figure}[H]
		\centering
		\includegraphics[width=1\linewidth,scale=5]{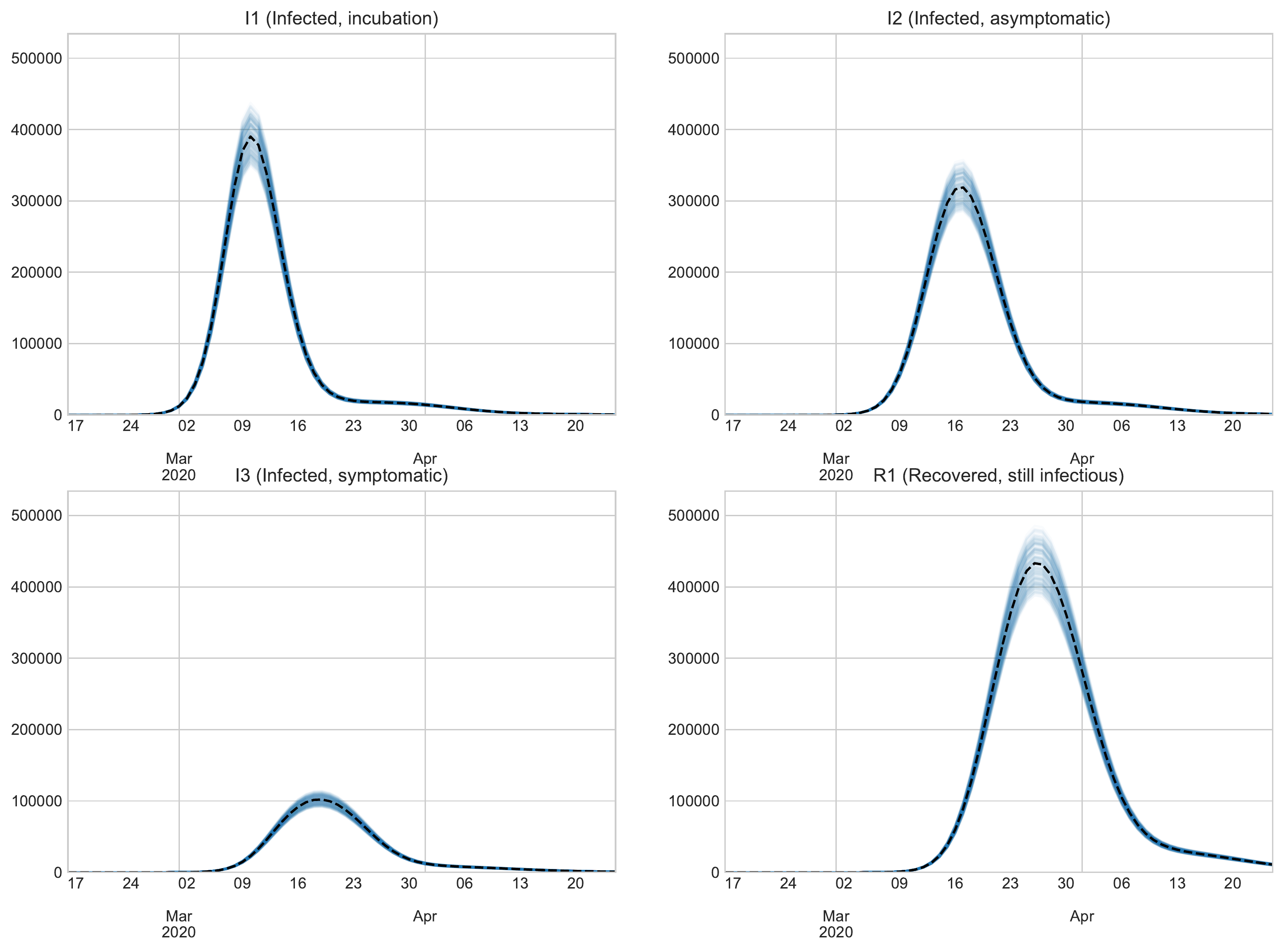}
				\includegraphics[width=1\linewidth,scale=5]{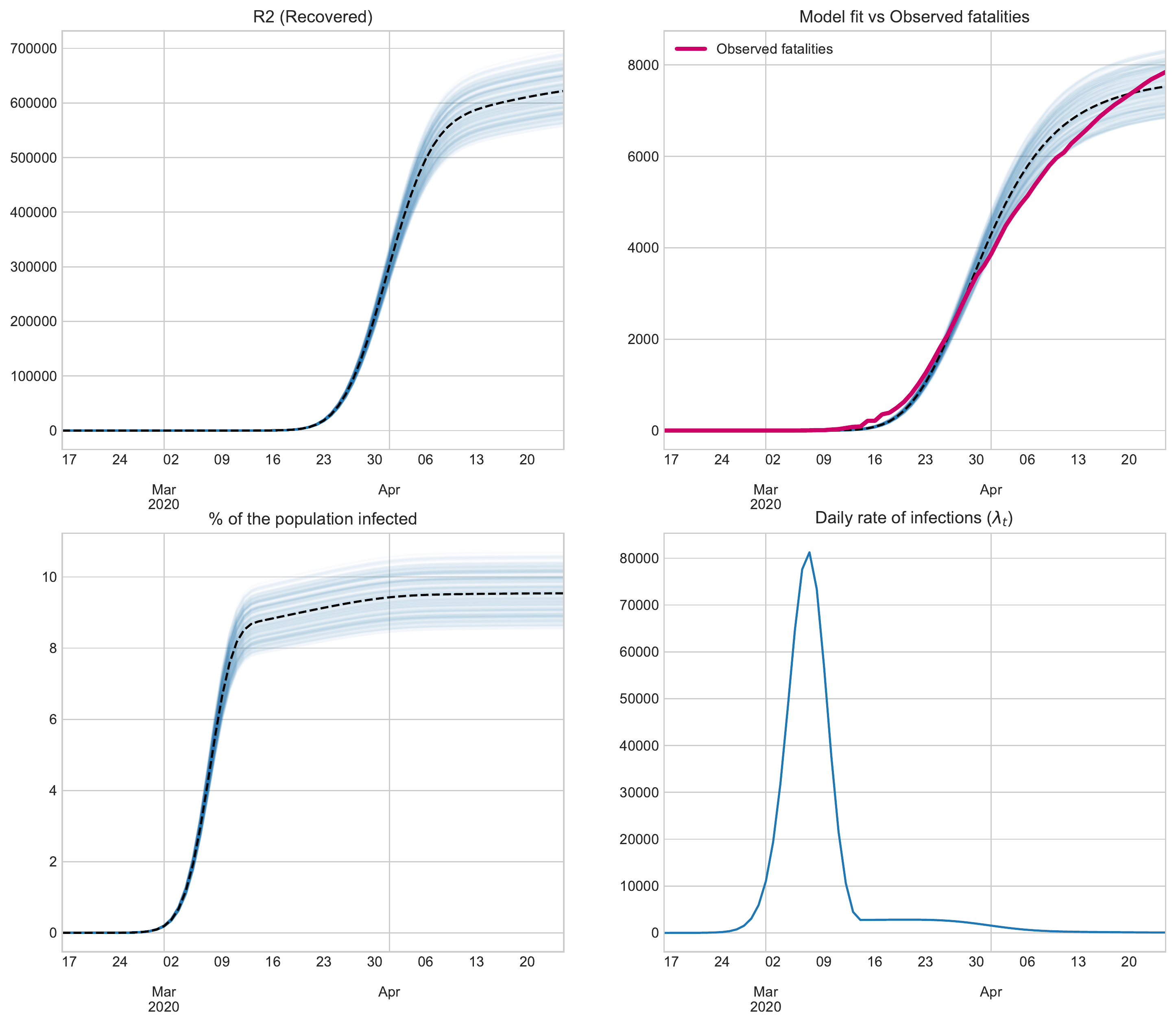}
		\caption{Results in Madrid: we assume that there are as many fatalities as in the official records.}
		\label{fig:fig1gal}
	\end{figure}

%

	\subsubsection*{Cataluña}
	
\begin{figure}[H]
	\centering
	\includegraphics[width=1\linewidth,scale=5]{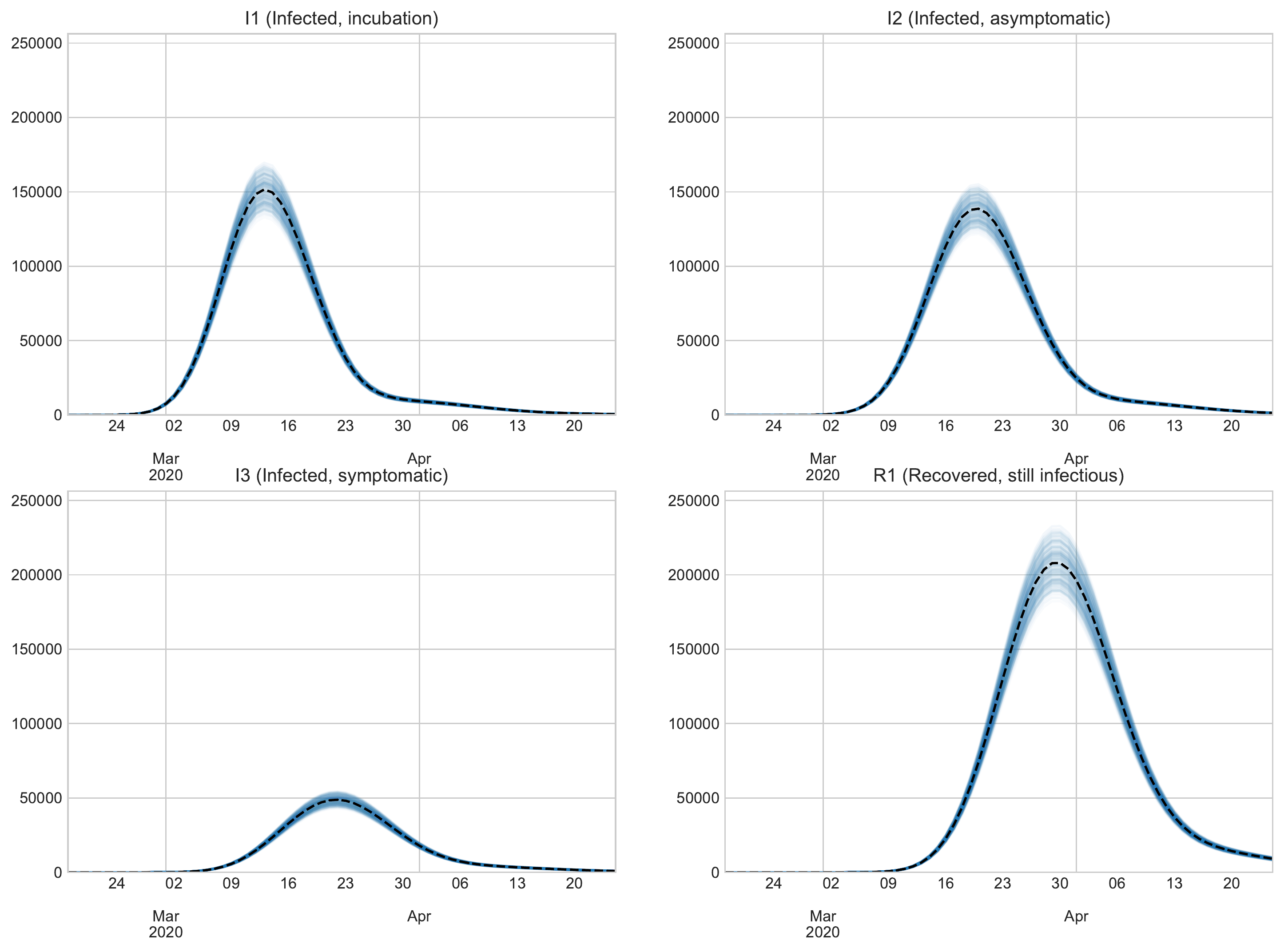}
		\includegraphics[width=1\linewidth,scale=5]{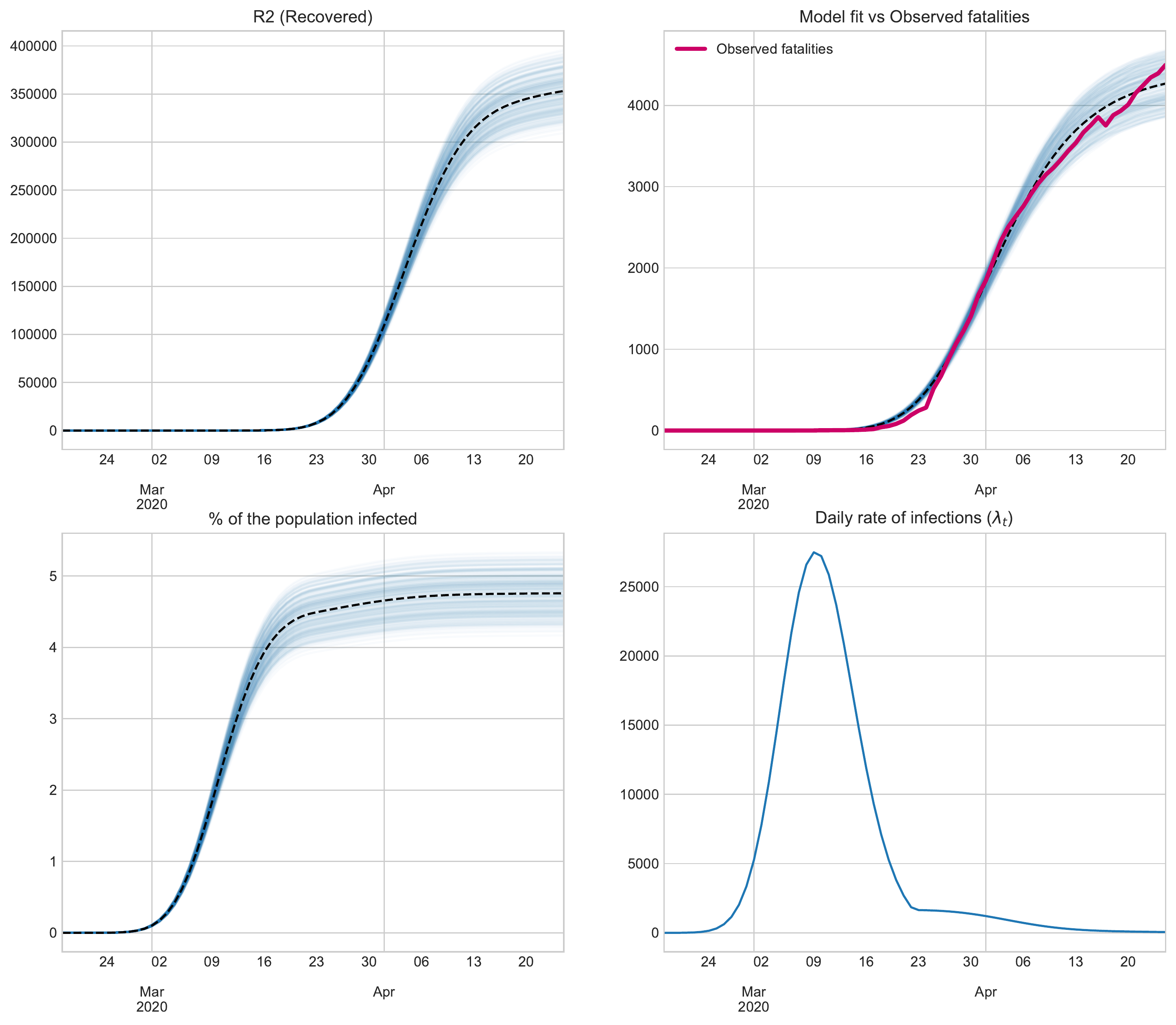}
	\caption{Results in Cataluña: we assume that there are as many fatalities as in the official records.}
	\label{fig:fig1gal}
\end{figure}


	\subsection{Scenario 2}
	
	\subsubsection*{Galicia}

	\begin{figure}[H]
		\centering
		\includegraphics[width=1\linewidth,scale=5]{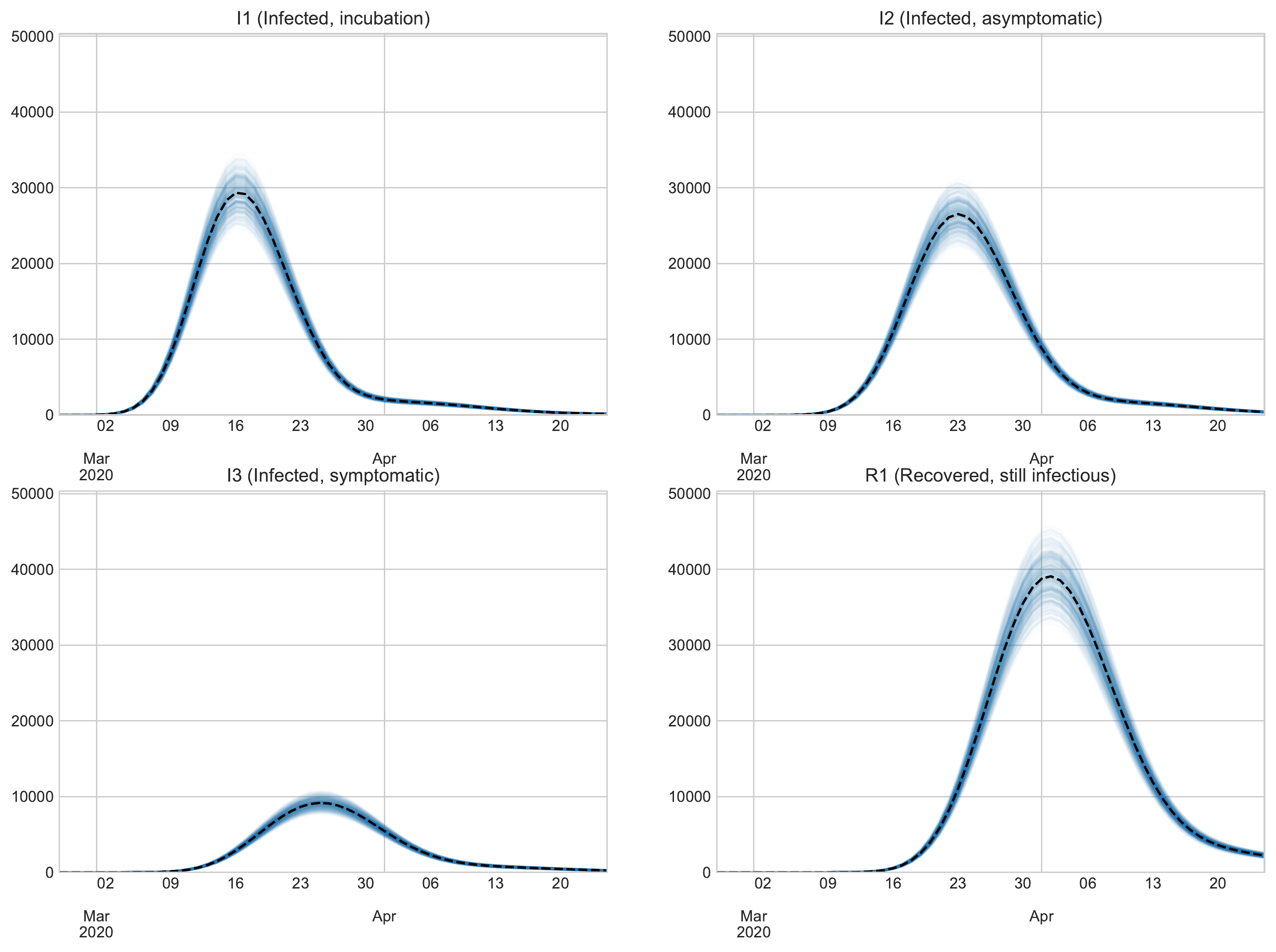}
				\includegraphics[width=1\linewidth,scale=5]{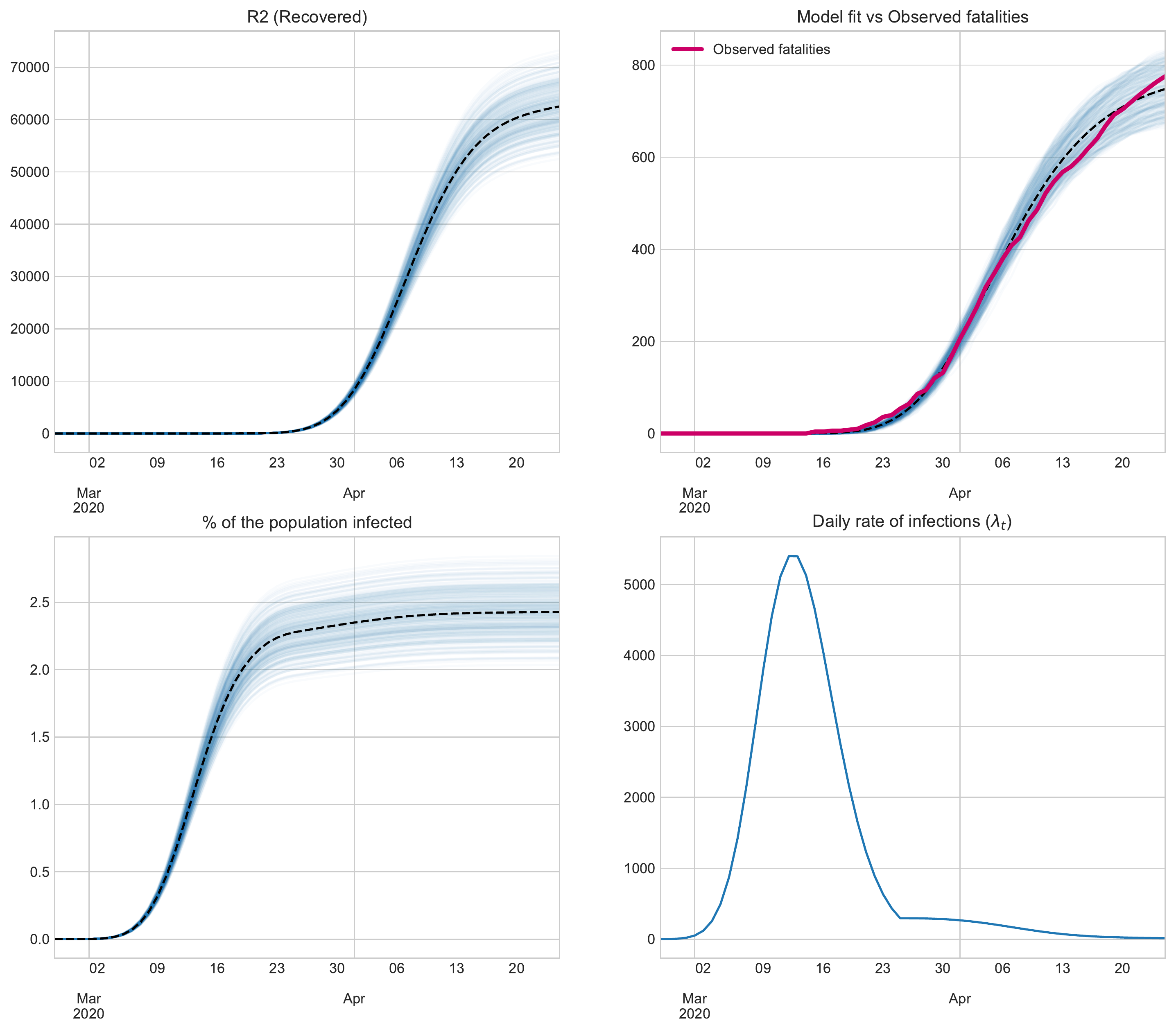}
		\caption{Results in Galicia: we assume there are twice as many deaths as the official records.}
		\label{fig:fig1gal}
	\end{figure}

%
%
	
	\subsubsection*{Castilla y Leon}
	
	\begin{figure}[H]
		\centering
		\includegraphics[width=1\linewidth,scale=5]{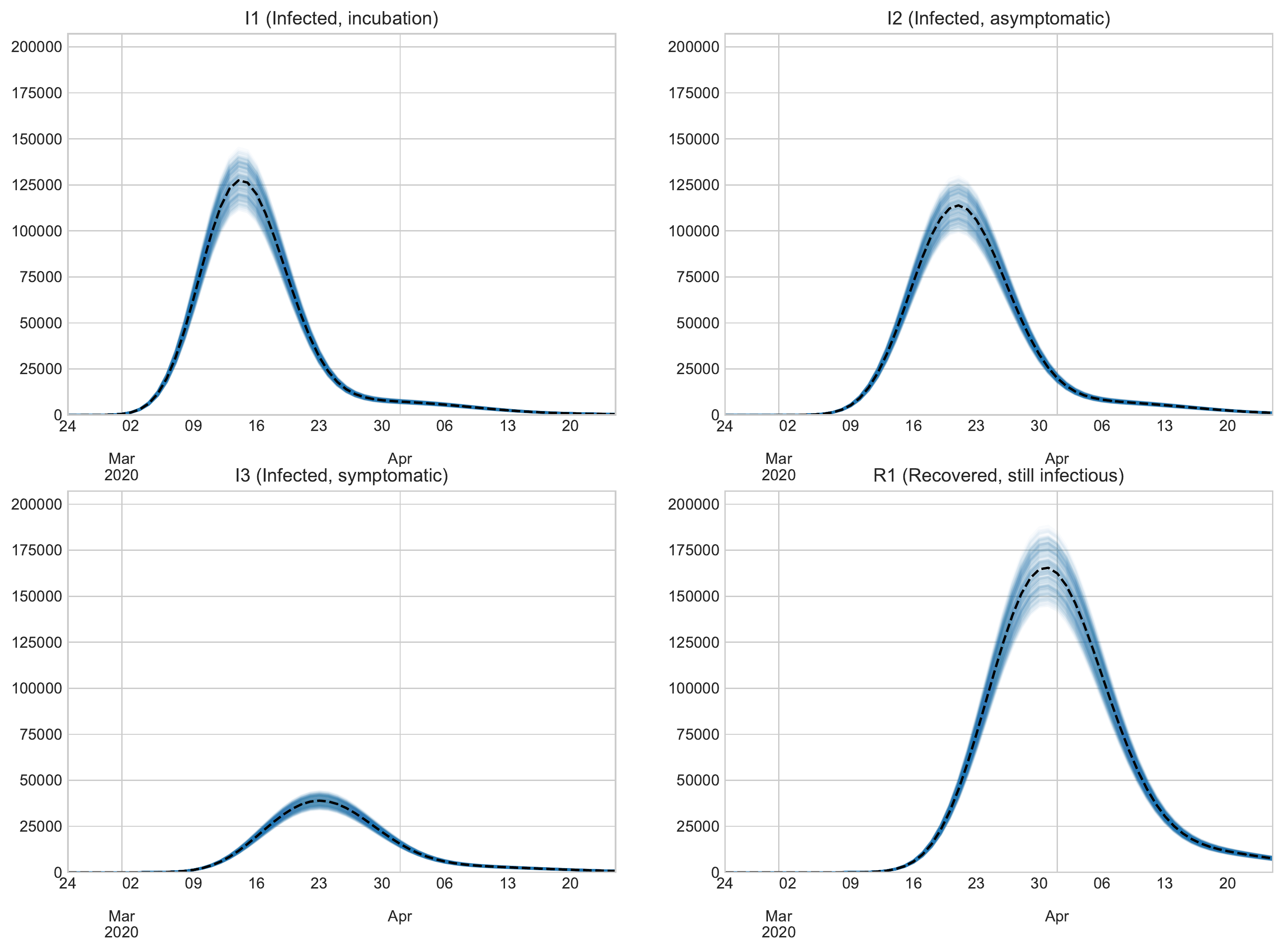}
				\includegraphics[width=1\linewidth,scale=5]{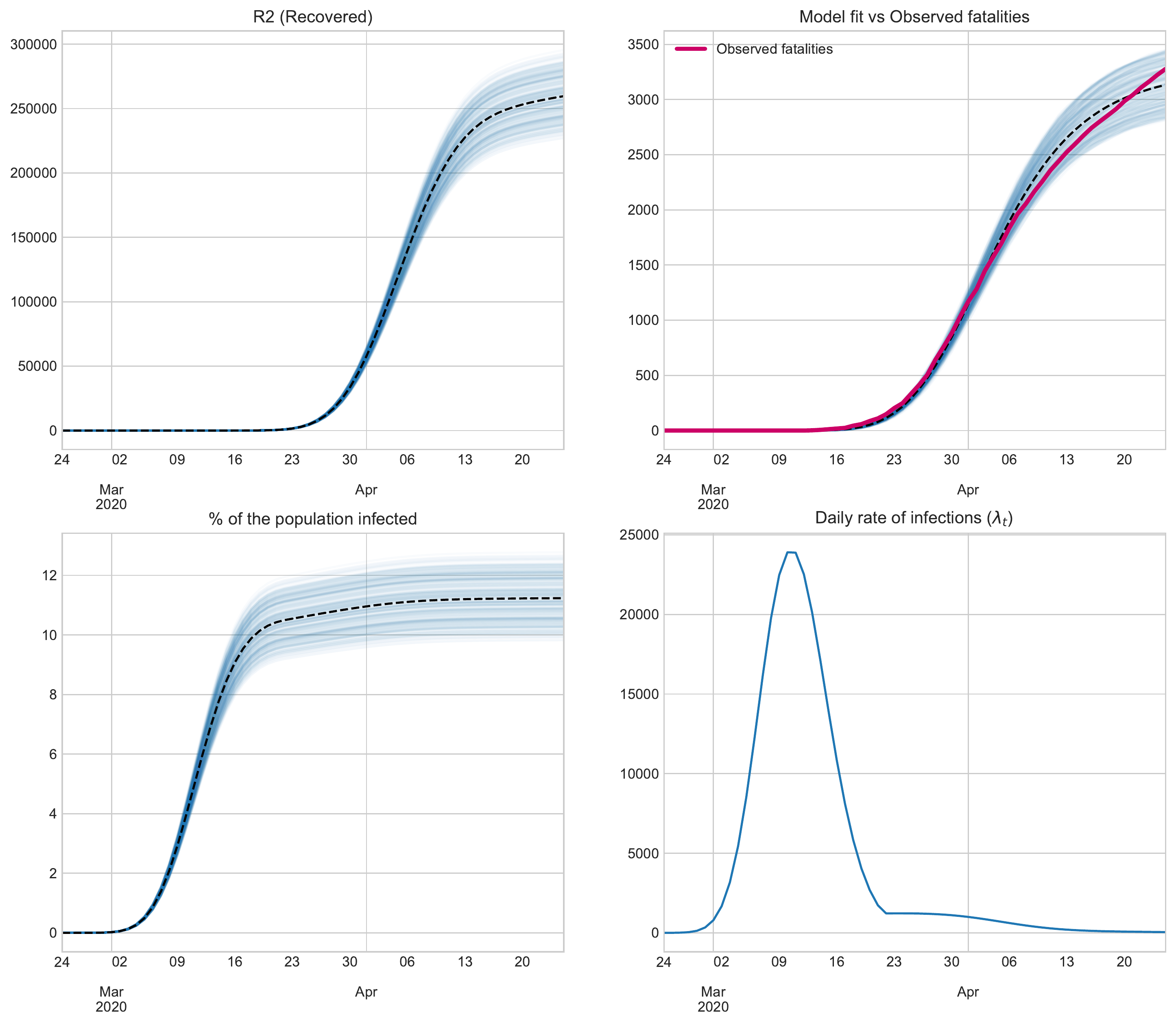}
		\caption{Results in Castilla y León: we assume there are twice as many deaths as the official records.}
		\label{fig:fig1gal}
	\end{figure}

%

	\subsubsection*{País Vasco}

	\begin{figure}[H]
		\centering
		\includegraphics[width=1\linewidth,scale=5]{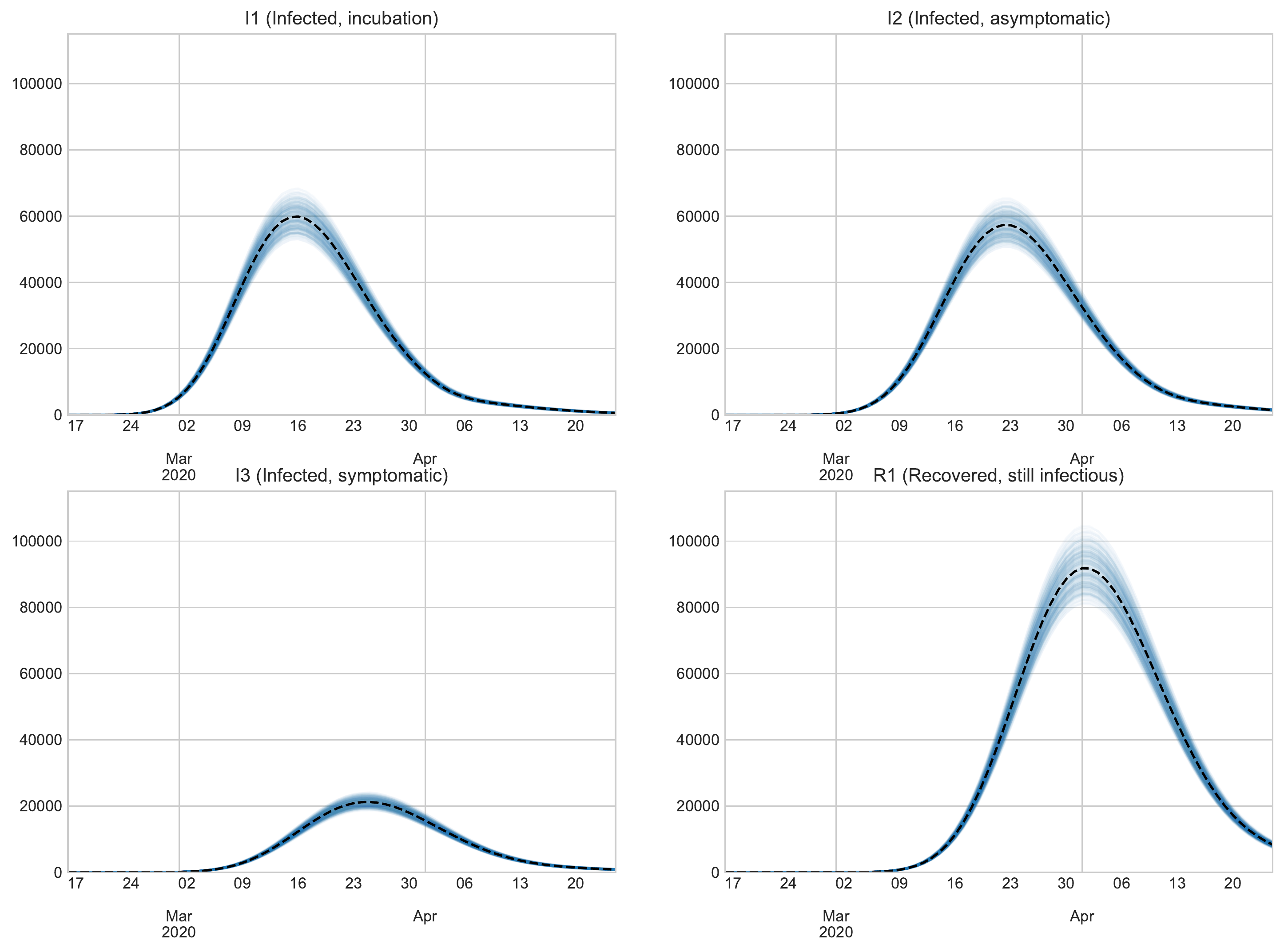}
				\includegraphics[width=1\linewidth,scale=5]{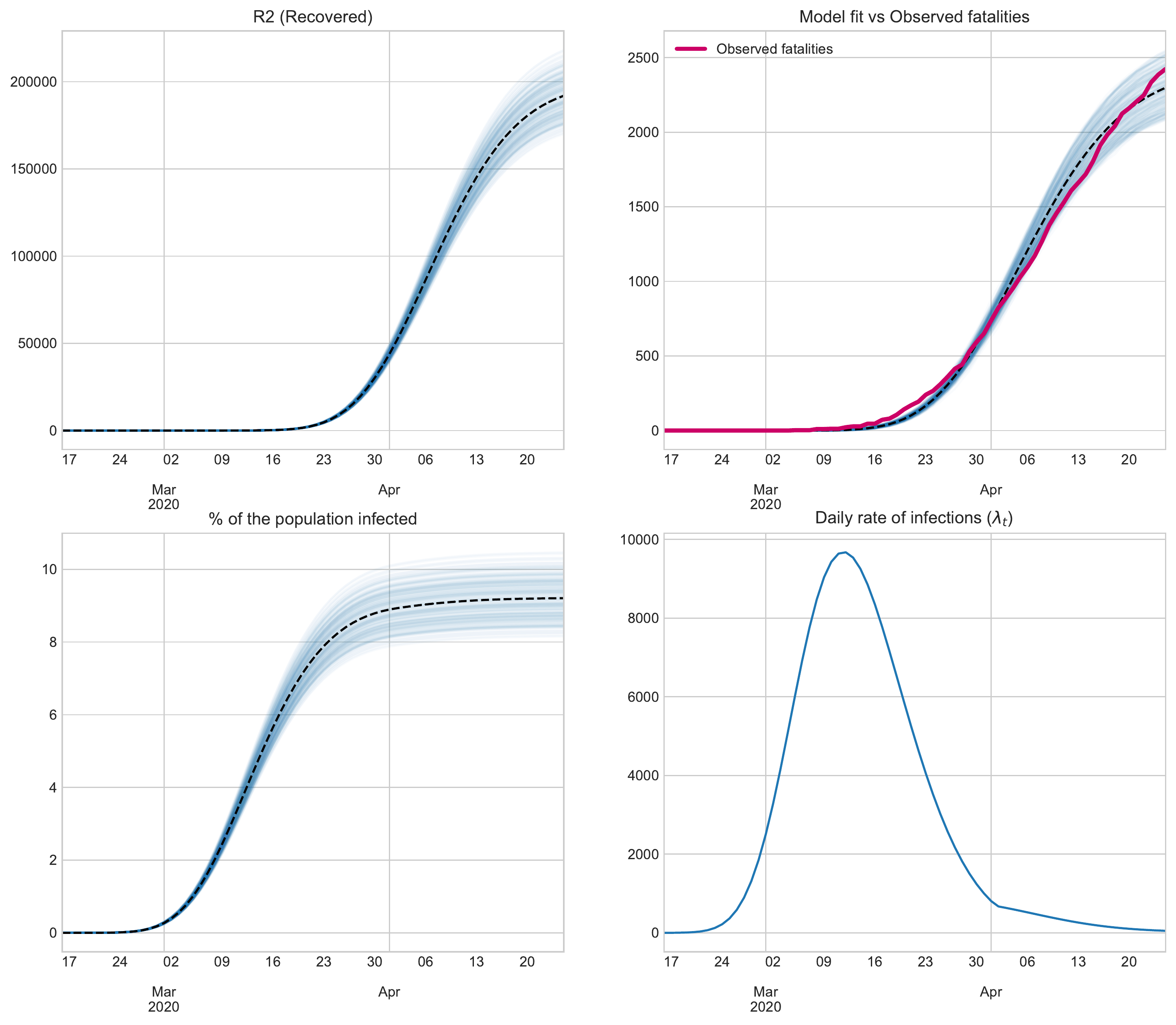}
		\caption{Results in País  Vasco: we assume there are twice as many deaths as the official records.}
		\label{fig:fig1gal}
	\end{figure}

%

	\subsubsection*{Madrid}
	
		\begin{figure}[H]
		\centering
		\includegraphics[width=1\linewidth,scale=5]{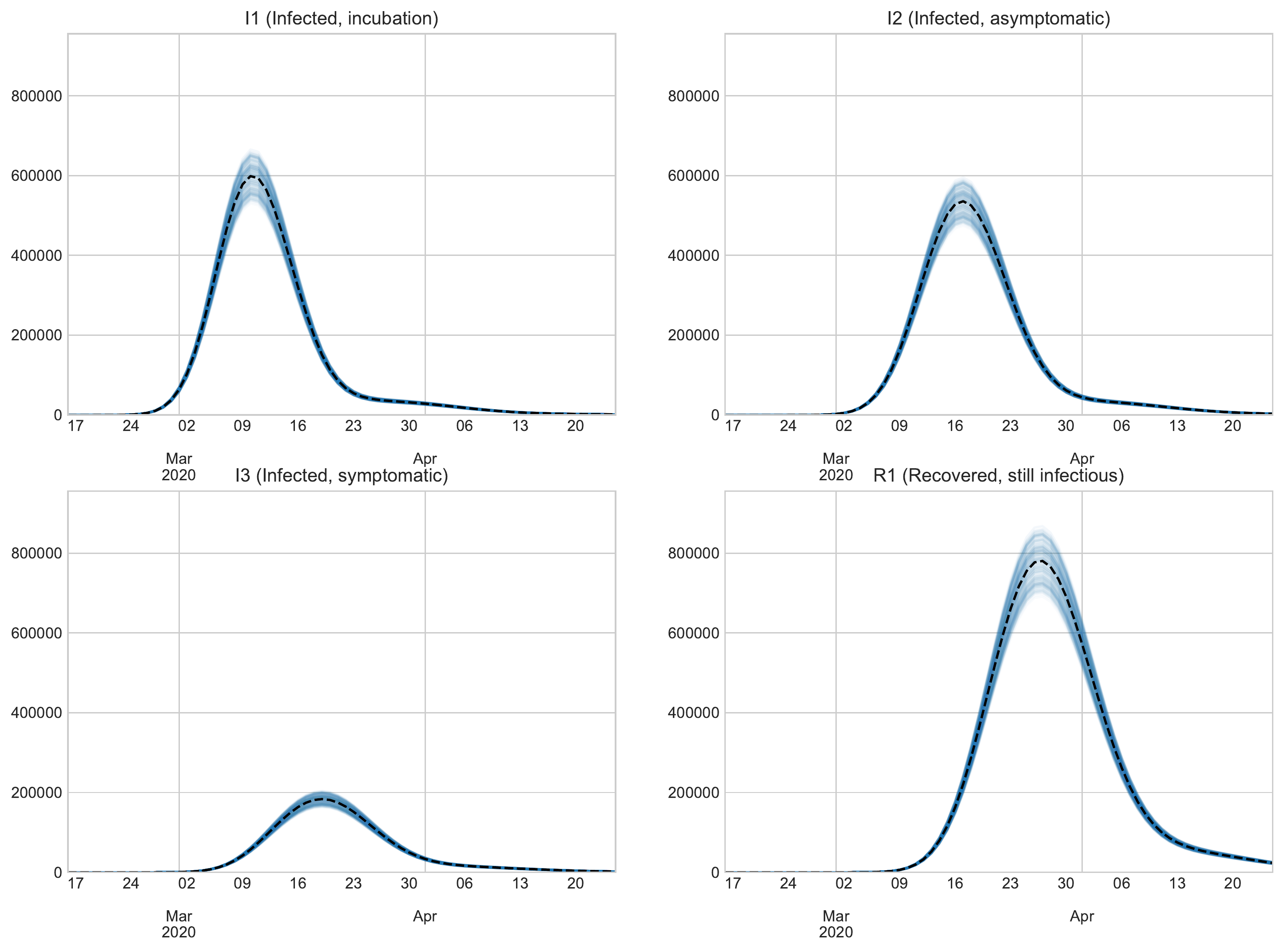}
				\includegraphics[width=1\linewidth,scale=5]{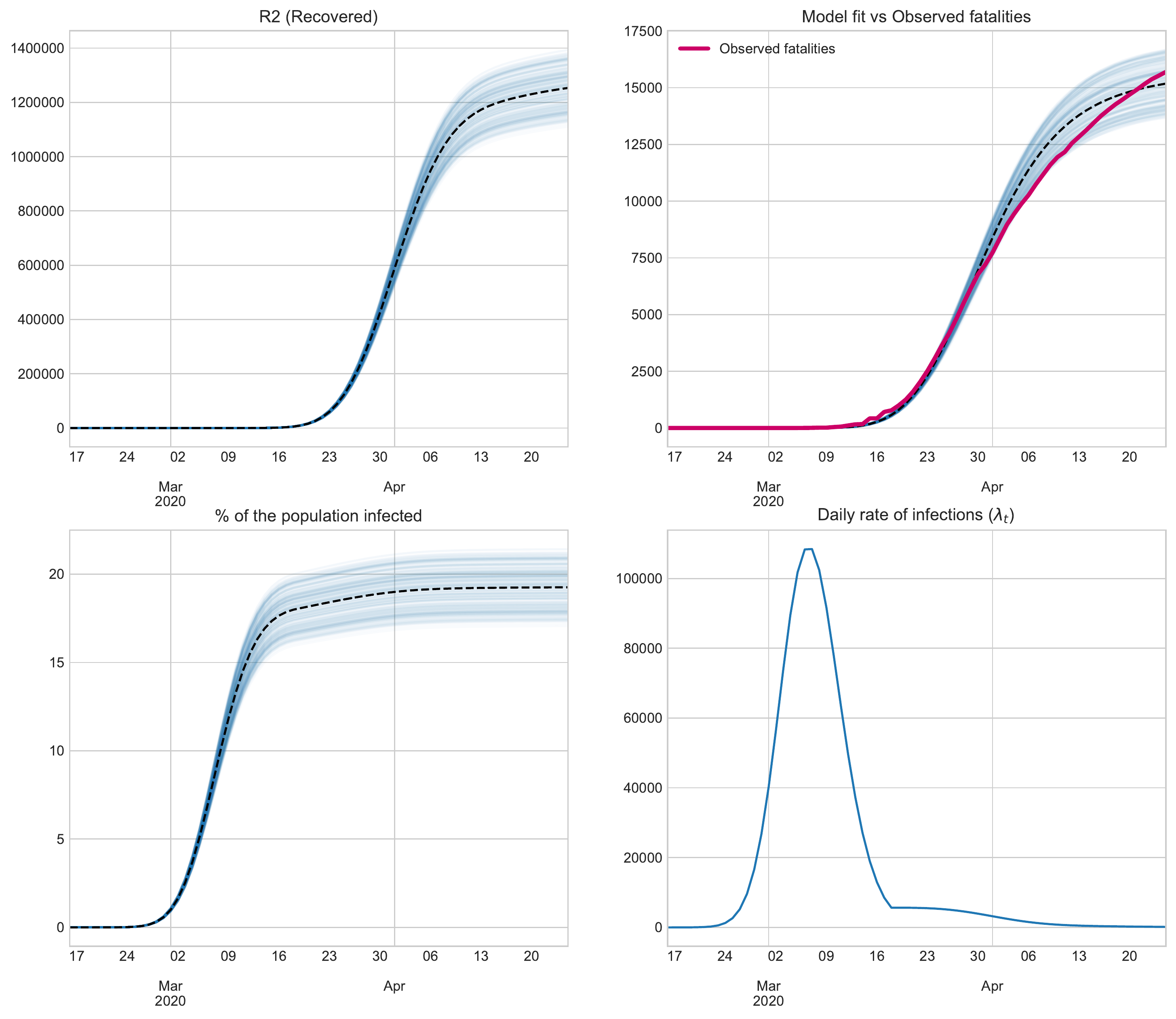}
		\caption{Results in Madrid: we assume there are twice as many deaths as the official statistics.}
		\label{fig:fig1gal}
	\end{figure}


	\subsubsection*{Cataluña}
	
	\begin{figure}[H]
		\centering
		\includegraphics[width=1\linewidth,scale=5]{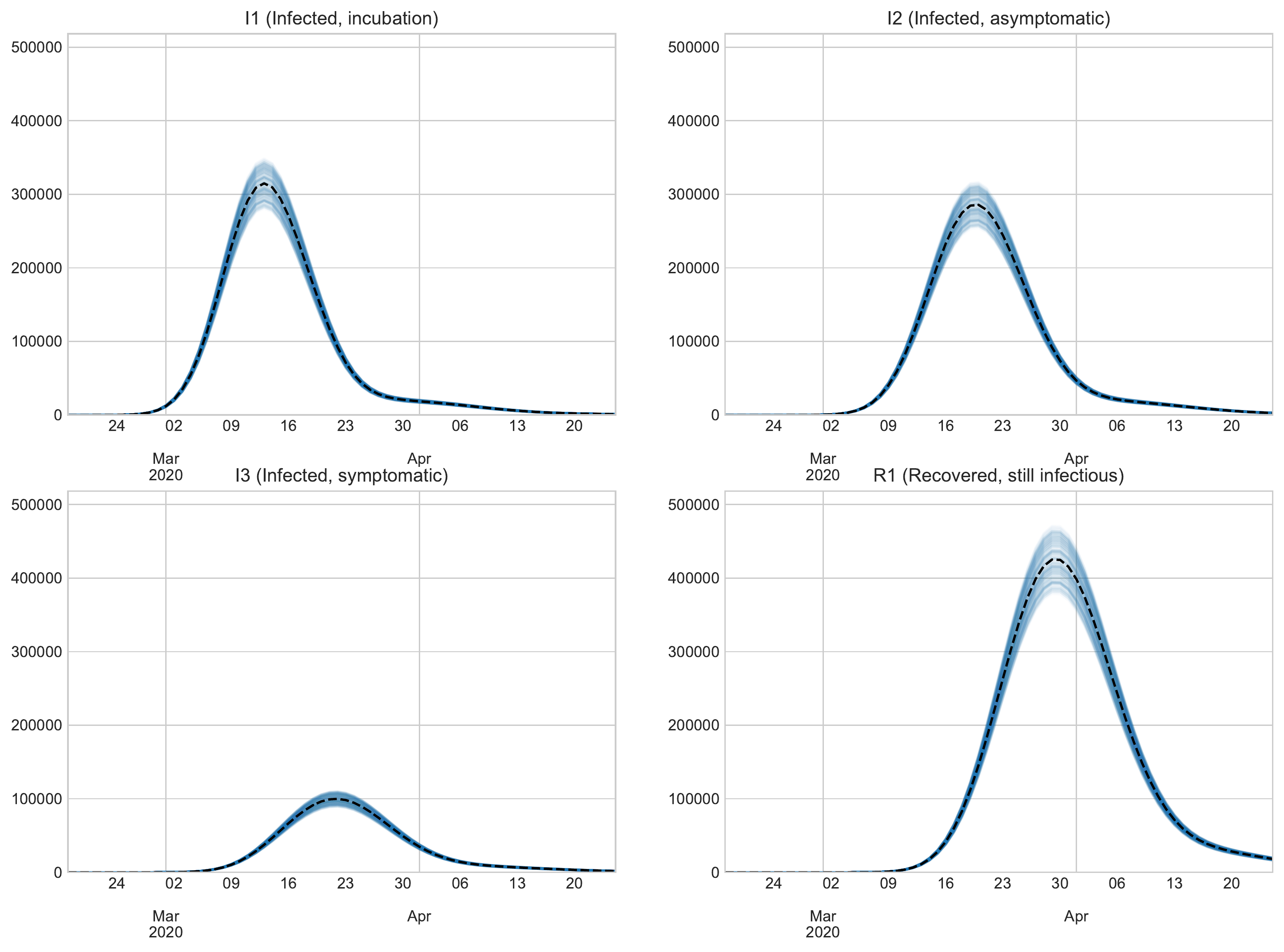}
				\includegraphics[width=1\linewidth,scale=5]{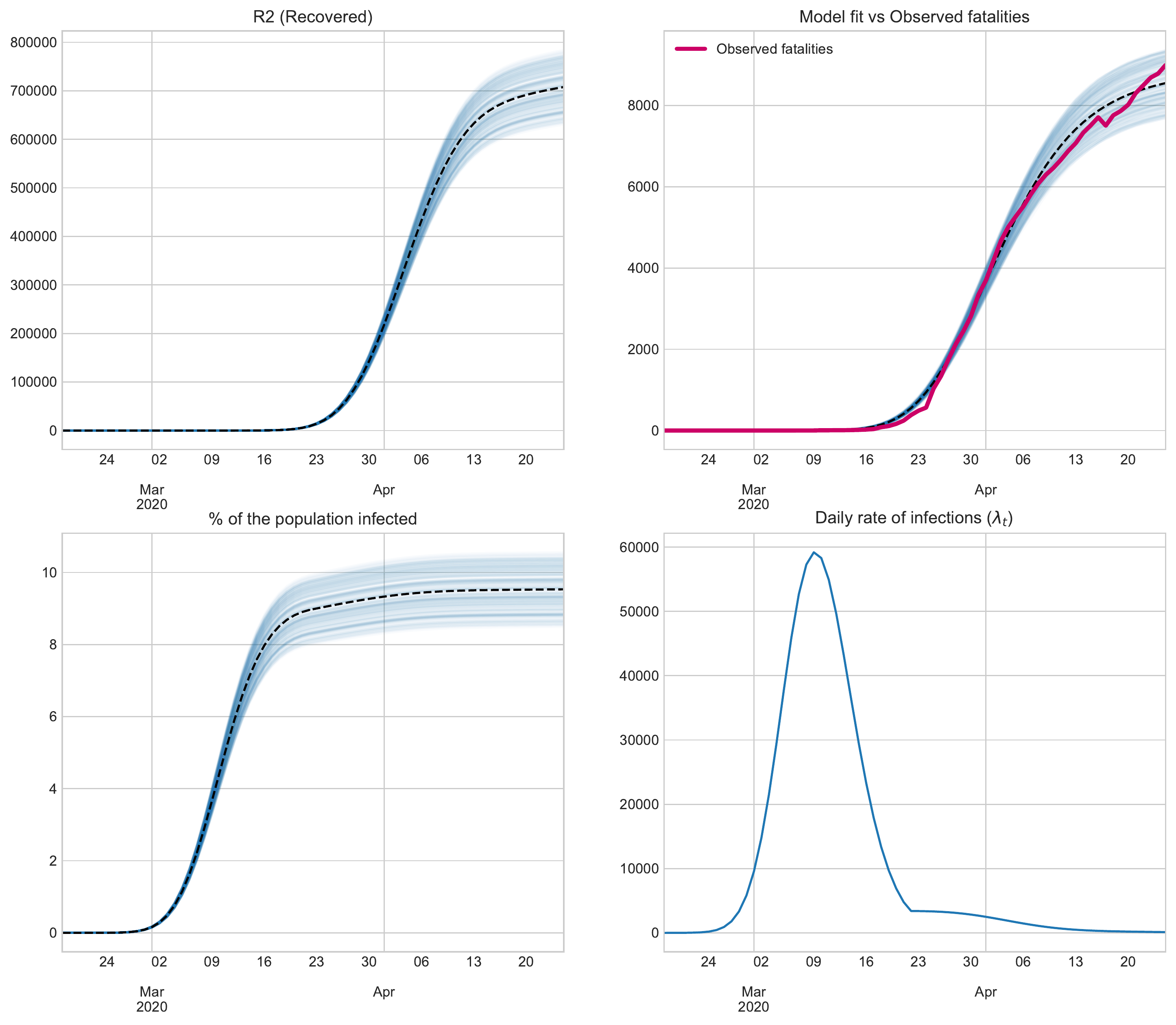}
		\caption{Results in Cataluña: we assume there are twice as many deaths as the official statistics.}
		\label{fig:fig1gal}
	\end{figure}

%

   \subsection{Analysis of results}

  The most relevant results are described below:
 
 \begin{itemize}
 	\item     Madrid is the most affected the region by COVID-19. $19.5$ \% of the population could have been infected or recovered from the virus if we assume that the number of deaths is double that reported by the Government. At present, there may be $120,000$ patients recovered.
 	
 	\item     Galicia is the region which suffered the mildest effects. The percentage of infected people is less than $2.5$ \%.
 	
 	\item     Castilla y León,  Pais Vasco, Cataluña could have suffered the effects of COVID-19 with a proportional magnitude. The percentage of infections could be between $6-10$ \% of the population.
 	
 	\item     The peak of new infections has probably occurred at the start of quarantine, while the peak of people who can contaminate took place between March 17-24.
 	
 	\item The number of new infections have been dramatically reduced after the introduction of containment measures. Now, it seems that the situation is back under control.
 	
 \end{itemize}

\newpage

\section{Discussion}

The Coronavirus pandemic is causing an unprecedented crisis in health, economic, and social terms throughout the world. To help designing  better politicies in the future, estimating the number of people who were infected by the virus is a crucial priority.  To address this issue,   we have proposed a new probabilistic model to estimate the spread of an epidemic by simulating the behaviour of each individual in a population.

\subsection*{COVID-19 in Spain}

The predictions obtained over several regions of Spain have practical relevance.   The results reveal that the number of infections  is much higher than what is reported by the Spanish government. Two factors may explain these discrepancies: i) The tests were carried out in a limited way among the general population; and ii) no random sampling was done among the different regions throughout the territory shading the real magnitude of the pandemic.  It is also worth noting that its impact was uneven across regions. Madrid was the most affected region, while Galicia was the least affected.  However, according to our model, in Cataluña, Castilla y León, and the País Vasco, the effects were surprisingly similar, even though the outbreak of infections theoretically started at different time points. In addition, the geographical dispersion of Castilla y León is more significant (see Table \ref{table:regiones}). Perhaps the older population of Castilla y León (see Table \ref{table:regiones}) has a higher case fatality rate than the other communities, and our model is overestimating the number of infected people in that region. As previously mentioned, taking into account the demographic structure of the regions, together with the epidemiological profiles and other socioeconomic characteristics, is fundamental in the analysis of results.

\subsubsection*{More deaths than those reported in the official records}

Even though many patients die due to the virus, this is not the official cause of death.  This problem occurs due to a lack of diagnostic tests. Spanish and International press echoed this problem in the current Coronavirus crisis, and more than twice as many deaths than those officially announced have been reported in certain regions. A recent study reported that the actual number of deaths might be even three times higher in Italy  \cite{buonanno2020news}.

Due to this fact, it is vital to take into account this uncertainty present in the models to perform an accurate estimate of the number of actual infections. However,  this uncertainty is not probabilistic.  To alleviate this problem, we have assumed that there could be up to twice the number of official deaths.

If we consider this correction, the number of infected increases from $40$ to $110$ percent by region.

\subsubsection*{Practical considerations of this results}

These findings allow us have a better sense of the degree  of exposure to the virus of the population and to design a restoration to normality based on the following factors.

\begin{itemize}
	\item The degree of immunity of each population.
	\item The number of people who might contaminate today.
	\item The number of recovered patients.
\end{itemize}

In this sense, it is vital to gradually de-escalate the confinement and always proceed according to the individual characteristics of each region.  

For the moment, there is no available vaccine for use with humans, and there is uncertainty about treatments. Therefore it is advisable to take measures such as social distancing \cite{Woody2020.04.16.20068163} to prevent the infection rate from skyrocketing and to avoid another outbreak.

\subsubsection*{The notorious peak}

Wondering about when the peak of infection would have been reached was a hot topic in mass media. However, such peak can be i) the maximum peak of new infections. ii) the maximum peak of people who can contaminate. In the first case, our models prove the existence of this peak between $9-16$ March, while in the second case, it was estimated to be in March $17-24$.

From a perspective of virus containment, the most important is the second one from as it indicates the number of potential transmitters.

	\subsection*{Methodological aspects of epidemic modeling}
	
	From a methodological standpoint, indeed, there are already simulators that work at the individual level \cite{Bisset2012,10.1371journalpcbi1000656,inproceedings,6063077,Mao2010}.   However, from our point of view, the models analyzed have some of the following potential limitations:   i) some models use fixed parameters without prior optimization according to the target population; ii) the parameters do not usually evolve dynamically according to the policies introduced; and iii), many of these models were designed without adding biological expert knowledge. 
	

	More generally we can find two philosophies when addressing epidemic forecasting  in the literature (see \cite{10.1371/journal.pcbi.1006134}): i) Mechanistic models: Using an epidemiological approach, these models explain the evolution of an epidemic from the causal mechanisms of transmissionc \cite{Koopman1999}; ii) Phenomenological models: From entirely data-driven approaches such as regression models or time series, the development of epidemics is predicted using historical data \cite{10.1371/journal.pcbi.1006134}.

Phenomenological models can be an appropriate solution in the case of epidemics such as influenza \cite{w4}, for which there exists long-term data collected on hospitalization and mortality rates. Additionally, the etiology, together with the previous incidence rates,   are better characterized. With all this, extrapolations of the real impact of the epidemic in the whole population can be done from data-driven approaches. In addition, current evidence indicates that influenza incubation times are shorter than those of COVID-19, which reduces an individual's potential exposure to the virus.   For all these reasons, we believe that in the case of the new COVID-19, it is necessary to use Mechanistic models as our proposal.

Often, defining a clear boundary between Mechanistic models and Phenomenological models may not be simple, and even both methods can be seen as complementary approaches. For example, SIR models and their variants from a theoretical point of view try to explain the effects of an epidemic from the mechanisms of transmission, however they are often used as time series \cite{bjornstad2002dynamics}. At this point, we would like to point out that the underpinning assumptions of these models are often unrealistic \cite{Huppert2013}. Moreover, from the perspective of Epistemology and Epidemiology, the epidemic spread is based on the complex interactions between individuals, which these models cannot capture due to their population-based nature \cite{10.1371/journal.pcbi.1006134}.

	The use of individual models should be an aspiration in the development of new methods.  In the current era of big data, in developed countries, a substantial part of the population is geolocalized through mobile devices. Moreover, it is increasingly common to use biosensors that monitor patient health \cite{Li2017}. Knowledge about interactions between individuals and patient health can provide a more realistic picture of the spread and evolution of the epidemic \cite{lorch2020spatiotemporal}. However, in this context, it is necessary to systematically apply techniques that correct the biases that appear in the observational nature of these data. At the same time, it is also important to preserve the privacy of the citizens.

The absence of reliable information is one of the main problems with the COVID-19 pandemic. To alleviate this problem, when fitting model parameters, it is important  to understand the social, economic, and demographic characteristics along with the epidemiological profiles of the region to introduce expert knowledge into the model and reduce uncertainty in statistical learning. Also, with this strategy, we reduce the computational demands and we increase model interpretability.

	\subsection*{Collaborative attitude }

	Information is one of the most valuable weapons in combating a pandemic of this magnitude. Understanding the mechanisms of virus transmission \cite{zhang2020evolving,Luo2020.03.24.20042606}, the prevalence of asymptomatic patients  in the population \cite{10.1001/jama.2020.2565},  risk and prognostic factors \cite{hernan} are  topics that can reduce the number of deaths. Due to the absence of prior knowledge of all these factors, a collaborative approach between countries and institutions to share their knowledge should be a top priority.

At an international level, we could highlight the initiative promoted by Eran Segal from Israel, to which several countries have joined \cite{Segal2020.04.02.20051284}. Segal and his team designed a survey in which anonymous volunteers participated to estimate the real evolution of the Coronavirus in Israel \cite{Rossman2020}. This questionnaire was adapted and applied in other countries. There is now an international consortium in action to share the results of these surveys, together with other clinical information from patients. However, it is important  to note that the data obtained from the questionnaires do not represent a random sample of the population, which limits the quality of the data collected.

	\subsection*{Expert knowledge in parameter fitting}
	
	From a general perspective of inverse problems and statistical learning, the estimation of the model parameters is an ill-posed problem \cite{o1986statistical,doi:10.1080/00029890.2001.11919778,bickel2006regularization,vapnik2013nature,vito2005learning,vapnik2015v}.  In the case of epidemic spread models, these difficulties are even more significant (see, for example, \cite{Xiang2015,Smirnova2017}).  In practice, the evolution of several compartments is predicted simultaneously, and usually, there is only information about the actual growth of one of the compartments.  
	
In this regard, it is vital to follow Ockham's Razor rule: the simplest solution is most likely the right. Reducing the number of variables to be fitted in the models and introducing expert knowledge into these models guided by current biological evidence is  for  ill-posed estimation problems.

In the case of COVID-19, the current epidemiological evidence is inconsistent. It is especially noteworthy about the case fatality rate (CFR)  and the proportion of asymptomatic patients \cite{doi:10.1056/NEJMe2002387}. In the first case, variations between $0.4$ and $15$ percent have been reported in the literature.   Meanwhile, with asymptomatic patients, these range from $20$ percent to $80$ percent. The primary rationale for these discrepancies is the application of non-observational data analysis methods for a sample composed largely of elderly patients. In this sense, after carrying out our estimates taking into account the demographic structure of Spain and using the published articles that we believe reflect the reality in a more precise way, we estimate that the rate of asymptomatic patients ranges between $75$ and $85$ percent. Likewise, CFR may vary between $0.4$ and $0.8$ percent. A recent article has estimated that asymptomatic patients to be $78$ percent \cite{Daym1375}.

	\subsection*{Combining a probabilistic model with a deterministic inverse problem}
	
	Our probabilistic model is non-stationary and has a complex dependency structure. In addition, the time series of each day's death records can be seen as a realization of the stochastic process. 
	
To fit the model, we estimate a deterministic inverse problem that makes the observed deaths is the average of the random process.  From there, we define a window, and we try with a non-probabilistic approach to capture the trajectories of fatalities close to the real one. In this way, we can obtain trajectories of the random process that allow us to explain the real situation of the infected in each region. In other words: From day one, with the chosen configuration of parameters, we can have many different scenarios in our probabilistic model; however, at the present moment, we already observed a concrete trajectory. Then, using this trajectory as a reference, we look for a set of close trajectories that measure other alternative scenarios that approximate the real number of deaths.

	\subsection*{Model limitations}

	The main limitation of our proposal is that some of the parameters on which the model depends on are set and may need to be tuned more carefully to the study population. However, the current biological evidence does not allow further refinement. Also, optimizing more parameters with a data-driven approach is dangerous due to the quality of data about infections reported by governments. Therefore, it is better to restrict oneself to current biological expertise information. 
	
	The models are an approximation of reality, and as Cox would say: all models are wrong, some are useful. We are predicting the unknown in this epidemic without reliable data. However, with statistical simulation models \cite{Amaran2014,Amaran2016}, we can see the effect of a deviation of the input parameters, the extreme cases, and the model predictions in other scenarios such as the end of containment policies.

	Another limitation of our model is that despite simulating data for each individual in the population, we were not using information about the demographic structure and epidemiological profile along regions. We also not introduce geolocation information as in the interaction between each individual in the population \cite{lorch2020spatiotemporal}. However, the inclusion of all these aspects, may not lead to greater accuracy as no reliable model output data are available,  and, furthermore, we would be entering into privacy issues that are beyond the scope of this study.
	
	\subsection*{Future work and model extensions}

Our immediate work is to extend our estimations to other countries. Also, we want to simulate the effect of an immediate end coronavirus lockdown \cite{Peak2020.03.05.20031088} in each of these location.  We also want to carry out simulations to find out how to minimize the effects of a possible outbreak in Autumn \cite{kissler2020projecting}, for example, with social distance \cite{Kissler2020.03.22.20041079} or weeks of  partitioned work: combining remote working with face-to-face activity. In a complementary way, it would be interesting to predict the effects of these measures on the economy. Some economists claim that the economic consequences of this crisis may be more damaging to our health in the long term than the effects caused directly by the virus \cite{McKee2020}.

From a methodological point of view, another exciting extension is to develop a mixed model or an empirical Bayes
framework (see an example of this methodology \cite{Liu2020.03.28.20044578}) to make a simultaneous estimation across many regions and to be able to make a statistical inference about the parameters of the models.  In this case, it may be interesting to optimize the parameters with Bayesian optimization \cite{shahriari2015taking}, to alleviate the computational issues of this approach. Another variation of the model may be to employ other probability distributions such as the Conway-Maxwell-Poisson distribution \cite{doi:10.1111/j.1467-9876.2005.00474.x} to handle underdispersion situations in the number of newly infected. However, our current election of  Poisson distribution is well  motivated given the asymptotic properties of the Erdos-Rényi model \cite{house2012modelling}.
	
	Finally, another possible improvement of the model could be to introduce an ensemble forecasting as they do in meteorology and manage better  the possible uncertainty in the initial conditions \cite{zhang1997ensemble}.

	\subsection*{Our future prediction in the world}
	
	Given the current emergency, and the need to provide predictions on the spread of the epidemic worldwide, we have decided to set up a website (\url{https://github.com/covid19-modeling/forecasts}) where we regularly update our estimates across many regions of the world in a fully accessible manner.  In addition to this, we will be incorporating our simulations of what the effect of current end coronavirus lockdown would be in each of these regions.

	\newpage

	\section{Mathematical model}
	
	\subsection{Model elements}

	We suppose that $D=\{0,1,\cdots,n\}$ is the set of days under study. Consider the following random processes whose domain is defined on D.
	
	\begin{itemize}
		\item  $S(t)$: Number of people susceptible to become infected on day $t$.  
		\item  $I_1(t)$: Number of infected individuals who are incubating the virus on day $t$.
	\item  $I_2(t)$:     Number of infected people who have passed the incubation period and do not show symptoms on day $t$.
	\item  $I_3(t)$:  Number of infected people who have passed the incubation period and do show symptoms on day $t$.
	\item  $R_1(t)$: Number of recovered cases which are still  able to infect on day $t$.
	\item  $R_2(t)$. Number of recovered cases which are not able to infect anymore on day $t$.
	\item  $M(t)$: Number of deaths on day $t$.       
	\end{itemize}

	Henceforth, we will denote by $I(t)= I_1(t)+I_{2}(t)+I_{3}(t)$ the number of infected people  at  time $t\in D$ and $R(t)= R_{1}(t)+R_{2}(t)$ the number of recovered people.

	The above random processes describe the evolution of population in different compartments. However, unlike the classics models \cite{keeling2011modeling,ball2019stochastic}, we divided the infected and recovered individuals in a more broader and realistic taxonomy for the particular case of the COVID-19.  There are two main reasons for this: First, the patients tested by healthcare are usually those found in $I_3$. In this case, there exists a corpus of prior knowledge about how they evolve over time and in case of death, which is their survival time. Second, there is evidence that there are recovered patients who can still infect others.

	\subsection{Model definition}
	\label{section:modelo}

	\begin{figure}[ht!] 
		\label{figura}
		
		\centering
		\begin{tikzpicture}
		\Vertex[x=-3,y=2]{S}
		\Vertex[x=0,y=2,L=$I_1$]{I1}
		\Vertex[x=-1,y=4,,L=$I_3$]{I3}
		\Vertex[x=0,y=6]{M}
		\Vertex[x=4,y=6,L=$R_2$]{R2}
		\Vertex[x=4,y=2,L=$I_2$]{I2}
		\Vertex[x=3,y=4,L=$R_1$]{R1}
		\tikzstyle{LabelStyle}=[sloped]
		\tikzstyle{EdgeStyle}=[post]
		\Edge[label=$\alpha$](I1)(I3)
		\Edge[label=$1-\beta$](I3)(R1)
		\Edge[label=$\beta$](I3)(M)
		\Edge[label=$1-\alpha$](I1)(I2)

		\tikzstyle{EdgeStyle}=[post, bend right]
		\Edge[](R1)(R2)
		\Edge[](I2)(R1)
		\Edge[](S)(I1)
		\tikzstyle{EdgeStyle}=[post, bend left]

		\end{tikzpicture}
		\caption{Diagram of state changes in our model.}
			\label{modelo}
	\end{figure}
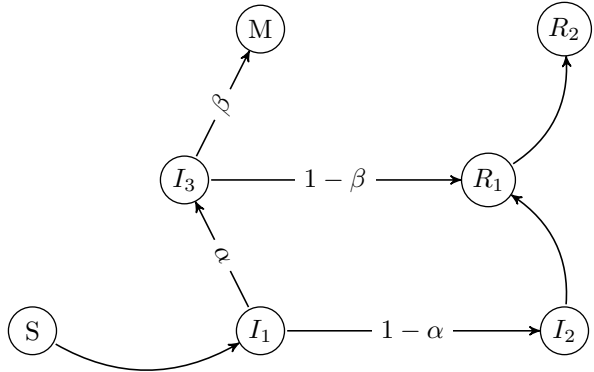

	The causal mechanism of new infected individuals  (see Figure \ref{modelo}) is introduced below. For each day $t\in D$,  we assume that the new infected population $I_{1}(t)-I_{1}(t-1)$, is generated by the individual interaction of the susceptible people with infected patients and the recovered cases who can still contaminate.  Formally, we assume that if a person can contaminate, it does so according to a random variable $X \sim \mathbb{\text{Poisson}} (R_i(t))$, being $R_{i}(t)$ the average number of new infections that can cause each person in the day $t$. It is important to remark, that it is natural to assume that the function $R_{i}(t)$ follows a decreasing trend over time in our setting, basically for two reasons: i) quarantine policies have been systematically introduced along with different countries and regions. ii) the number of susceptible people decreases over time, while the number of infected people can increase. This all makes it more complicated to interact with a non-infected people.

	Once a new infected person arrives to the model (see Figure \ref{modelo}), we assume that the transitions between the different graph states are modeled by a probability law that verifies the following conditions: i) the transition probabilities are independent of the absolute instant when such transition takes place ii) the probabilities depend only on the current state of the patient regardless of the previous path in the graph. In particular, given the $States=\{I_1,I_{2},I_{3}, R_{1},R_{2},M\}$, and $\alpha,\beta \in [0,1]$, we have: $\mathbb{P}(I_2|I_1)=\alpha$, $\mathbb{P}(I_3|I_1)=1-\alpha$, $\mathbb{P}(M|I_3)=\beta$, $\mathbb{P}(R_1|I_3)=1-\beta$, $\mathbb{P}(I_3|I_2)=1$, $\mathbb{P}(R_2|R_1)=1$; all other transitions take a value equal to zero in probability. More schematically, the transition matrix between events is shown in the Equation \ref{eqn:transicion}.

	\begin{equation}
	\label{eqn:transicion}
	P=
	\begin{bmatrix}
	
	& $I1$ & $I2$ &  $I3$ & $R1$  &  $M$ &  $R2$   \\
	$I1$ &    0 & 1-\alpha & \alpha & 0 & 0 & 0    \\
	$I2$ &    0 & 0 & 0 & 1 & 0  & 0 \\
	
	$I3$ &            0& 0 & 0 & 1-\beta & \beta &0 \\
	
	$R1$ &    0 & 0 & 0 & 0 & 0  & 1 \\
	$M$ &    0    & 0 & 0 & 0      & 1  & 0 & \\
	$R2$ &    0             & 0 & 0 & 0 & 0 & 1 \\

	\end{bmatrix}.
	\end{equation}

Additionally, the Table \ref{table:tabla2}  shows the random variables that model the time between transitions together with the references that have been used. In particular cases, as survival time, we have made our estimates based on data from some of these researchers and incorporating some expert knowledge of others.

	\begin{table}[ht!]
		\begin{center}

			\begin{tabular}{|c|c|c|}
				
				\hline 
				Transition    & Random variable & Used references  \\ 
				\hline 
				$I_{1}\to I_{2}$    & $Gamma(5.807,0.948)$ & \cite{lauer2020,abdel2020modeling}  \\ 
				\hline 
				$I_{1}\to I_{3}$    &  $Gamma(5.807,0.948)$ & \cite{lauer2020,abdel2020modeling} \\ 
				\hline 
				$I_{2}\to R_{1}$    &  $Uniform(5,10)$  &  \\ 
				\hline 
				$I_{3}\to R_{1}$    &  $Uniform(9,14)$   & \cite{abdel2020modeling} \\ 
				\hline 
				$I_{3}\to M$    & $Gamma(6.67,2.55)$   &  \cite{novel2020,surv1,abdel2020modeling,salje2020estimating}\\ 
				\hline 
				$R_1\to R_2$    & $Uniform(7,14)$ & \cite{bi2020epidemiology,ehmannvirological}  \\ 
				\hline
			\end{tabular}
		\end{center}
		\caption{Random variables of the time of each transition} 
					\label{table:tabla2}
	\end{table}

	Finally, in a similar way, in Table \ref{table:tabla3}, we show the values used to model the transition probabilities.

	\begin{table}[ht!]
		\begin{center}

			\begin{tabular}{|c|c|c|}
				
				\hline 
				Coefficient    &  Value & Used references  \\ 
				\hline 
				$\alpha$ & $0.8$  &  \cite{Daym1375,Nishiura2020.02.03.20020248,Tabata2020.03.18.20038125,:/content/10.2807/1560-7917.ES.2020.25.10.2000180}  \\ 
				\hline 
				$\beta$    & $0.06$  & \cite{rajgor2020many,doi:10.1056/NEJMe2002387,ref1,Wu2020,Mahase2020,dudel2020monitoring} \\ 
				\hline  
				
			\end{tabular}
		\end{center}
		\caption{Probability of each transition} 
					\label{table:tabla3}
	\end{table}

	\subsection{Model implementation}

	The proposed model is not known to have an closed-form solution. In a real-world setting, it is necessary to use statistical simulation methods to approximate the mean trajectory or quantile functions over time. Also, we must fit some parameters of the model to characterize the behavior of the study population. For this purpose, we use a sample of the deceased patients $M_1$, $M_2$,$\cdots$, $M_s$ of each day in the  timeframe $O=\{1,\cdots,s\}$.

	Next, we suppose that our model ($M$) is dependent on a vector of parameters $\theta= (\theta_1,\theta_2)\in \mathbb{R}^{p_1}\times \mathbb{R}^{p_2}$ (with $p_1+p_2=p$), where $\theta_1$ is a vector of dimension $p_1$, defined in beforehand, and $\theta_2$ must be estimated from the sample.  Furthermore, let us assume that the initial state of the system is characterized by  $S= (S(0),I_{1}(0),I_{2}(0),I_{3}(0),R_{1}(0),R_{2}(0),M(0))\in \mathbb{N}^{7}$ and $T= (T_1(0),T_2(0),T_3(0),T_4(0),T_5(0),T_6(0))\in \mathbb{N}^{m}\times \cdots \times \mathbb{N}^{m}$. $S$ has the number of elements for each compartment of the model on day $0$. $T$ also contains the amount of remaining days  to complete the transition they are in for each individual in the initial state, being $m$ a integer number that represents the maximun numbers of registred days.

	To simplify the notation, from now on we denote the average dead trajectory by the function $Mean(\theta_1,\theta_2,S,T)(t)$ for each day $t \in D$

	The next step is to estimate  $\hat{\theta}_2$.  To do this, we solve the following optimization problem:

	\begin{equation}  \label{eqn:1}
	\hat{\theta}_2= \arg \min_{\theta_2\in S\subset \mathbb{R}^{p_2}}\sum_{i=1}^{s} \omega^i (M_i- Mean(    \theta_1,\theta_2,S,T)(i))^{2},  
	\end{equation}
	
	where $\omega= (\omega^{1},\cdots,\omega^{s})$ is a weighted vector that can help to improve model estimation. Examples of this weights may be:
	
	$\omega^{i}= M_i/\sum_{i=1}^{s}M_i$ or  $\omega^{i}= (1/M_i)/(\sum_{i=1}^{s} 1/M_i)$ $(i=1,\dots,s)$.

	At this point, it is relevant to note that the above optimization problem (\ref{eqn:1}), as formulated, includes the possibility of introducing constraints in the space of parameters. It is important because we
	have prior knowledge of what the range of the parameters is. We
	may know constraints involving them as well.

	In Equation (\ref{eqn:1}), we have used the real mean trajectory. However, in practice, this is unknown, and we must approximate it using simulation.  Next,  we run $B$ differents simulations and we denote by $\overline{M}(\theta_1,\theta_2^{0},S,T) = \frac{1}{B} \sum_{i=1}^{B} M^{i}(\theta_1,\theta_2,S,T)$, the estimated mean trajectory.  $M^{i}(\theta,\theta_2,S,T)$ $(i=1,\cdots,B)$ denote the result of each simulation.

	So the optimization problem to be solved is:
	
	\begin{equation}  \label{eqn:2}
	\hat{\theta}_2= \arg \min_{\theta_2\in S\subset \mathbb{R}^{p_2}}\sum_{i=1}^{s} \omega^i (M_i- \overline{M}(\theta_1,\theta_2,S,T)(i))^{2}.  
	\end{equation}
	
	Schematically, the overall optimization process is described below.

	\begin{enumerate}
		\item Define an initial $\theta_2^{0}$ and run $B$ times $M(\theta_1,\theta_2,S,T)$. We denote by $M^{1}(\theta_1,\theta_2^{0},S,T)$, $\cdots$, $M^{B}(\theta_1,\theta_2^{0},S,T)$ to each of the obtained results.
		\item Estimate the mean trajectory $\overline{M}(\theta_1,\theta_2^{0},S,T) = \frac{1}{B} \sum_{i=1}^{B} M^{i}(\theta_1,\theta_2^{0},S,T)$
		\item Estimate the mean square error $\hat{RSS}^{0}= \sum_{i=1}^{s}\omega^{i} (\overline{M}(\theta_1,\theta_2^{0},S,T)(i)-M_i)^{2}$.
		\item To construct a succession of vectors $\{\theta_2^{j}\}^{M+1}_{j=1}$ so that $\hat{RSS}^{0}$ $>$ $\hat{RSS}^{1}$ $>$ $\hat{RSS}^{2}$ $>$ $\cdots$ $>$ $\hat{RSS}^{M+1}$. For example with a  stochastic optimization algorithm.
		
		\item Stop after $M+1$ iterations and return $\theta_2^{M+1}$ as the optimal parameter of the problem.
		
	\end{enumerate}

	In our particular setting, $\theta_2$ contain the parameter of a $R_{i}(t)$ function defines in Section \ref{section:modelo}. From now on, we asume that $R_i(t)= \min\{C,ae^{-(bt+ct^2+dt^3+et^{4}+ft^{5})}\}$ where $a\in [0,3]$, $b\in[-1,1]$, $c\in [0,1]$, $d\in [0,1]$, $e\in [0,1]$, $f\in [0,1]$ with $\theta_2= (a,b,c,d,e,f)\in [0,3] \times [-1,1]\times[0,1]\times[0,1] \times[0,1]\times[0,1]$ and $C$ is a positive constant fixed $0.005$.

\subsection{Model optimization}

In our setting, we have found optimal parameters taking into account randomness in the approximation of the mean. To do this, we must resort to stochastic optimization algorithms.

Many algorithms of this kind can be found in literature, but based on the good results obtained in a preeliminary analysis we have decided to use a state-of-the-art evolutionary algorithm: the CMA-ES \cite{hansen2016cma}.  CMA-ES is an evolutionary-based derivative-free optimization technique that can optimize a wide variety of functions, including noisy functions as the one we use in our method. One survey of Black-Box optimizations found it outranked 31 other optimization algorithms, performing especially keen on ``difficult functions'' or larger dimensional search spaces \cite{10114518307611830790}. From a theoretical point of view, CMA-ES can be seen as a particular case of Expectation-Maximization algorithm (EM) \cite{brookes2019view}.

    \subsection{Model Inference}
    
      Our model has been fitted independently for each region. Therefore, we cannot make statistical inference in the usual sense since we only observe a trajectory of a stochastic process with a complex dependence structure. Hence, to identify the model, we assume that the observed  trajectory of accumulated deaths is the mean of our random process.

    Given a day $t^{\prime}\in O$, $M_{t^{\prime}}$ accumulated number of observed deaths in $t^{\prime}$  and the multivariate random process \\ $\{X(t)=(I_1(t),I_2(t),I_3(t),R_1(t),R_2(t),M(t))\}_{t\in D} $, the death trajectories of the probabilistic model close to the observed path on $t^{\prime}$, are for us a set of possible explanations of the spread of the coronavirus in a study region. We now formalize this idea.  Let $I_{\alpha}=[M_{t^{\prime}}-\alpha M_{t^{\prime}},M_{t^{\prime}}+\alpha M_{t^{\prime}}]$, where $\alpha\in (0,1)$.  Considerer, $\{\omega\in \Omega: M(t^{\prime},\omega)\in I_{\alpha}\}$, being $\Omega$ sample space of process $\{X(t)\}_{t\in D}$. We build our confidence bands of level $\alpha$ as paths that belong $\{\omega\in \Omega: M(t^{\prime},\omega)\in I_ {\alpha}\}$.

    Our confidence bands contain the trajectories of infected and recovered patients that generate many deaths similar to that which occurs in reality. Therefore, in practice, $\alpha=0.1$ is usually a fair value, and $t^{\prime}$ is selected as one of the latest datums in the historical records.  

\subsection{Software details and resources}

Our proposal has been implemented in several programming languages: C++, Python, and R, although the results shown in this article have been obtained with Python. We optimized the parameters using library \verb| pycma| \cite{niko_2020_3764210}, and \verb|Numpy| has been used for mathematical operations.

In the different statistical analyses performed, we have used R. Plots have been made both in R with \verb|ggplot2| library and in Python with \verb|Matplotlib|.

Finally, the training data used to fit the models can be downloaded from at \cite{w5} and \cite{w6}.

In the immediate future we are going to release the code used in this paper  for the benefit of the scientific community at  (\url{https://github.com/covid19-modeling}).

\section*{Acknowledgements}

The authors thank Oscar Hernan Madrid Padilla for their helpful suggestions. We are also grateful to Felipe Muñoz López for his help in the computer implementations at the early stages of this research.

 This work has received financial support from the Consellería de Cultura, Educación e Ordenación Universitaria (accreditation 2019-2022 ED431G-2019/04) and the European Regional Development Fund (ERDF), which acknowledges the CiTIUS-Research Center in Intelligent Technologies of the University of Santiago de Compostela as a Research Center of the Galician University System.

\section*{Competing Interests}
The authors declare no competing interests.

\bibliographystyle{unsrt}
\bibliography{bibliografia}

\end{document}